\documentclass[pra, twocolumn,showpacs,nofootinbibfloatfix,amsmath,amsfonts,amssymb,table,xcdraw]{revtex4-1}%
\usepackage{amsmath,amsfonts,amssymb,color}
\usepackage{amsthm}
\usepackage{leftidx}
\usepackage{graphicx}
\usepackage{xcolor}
\usepackage{dcolumn}
\usepackage{bm}
\usepackage{epstopdf}
\usepackage{epsfig}
\usepackage{mathdots} 

\def\CH{\textcolor{black}}
\usepackage{comment}
\usepackage{verbatim}

\usepackage{caption}
\usepackage{graphicx}
\usepackage{subcaption}
\usepackage{braket}
\usepackage{amssymb}
\usepackage{mathrsfs}
\usepackage{cleveref}
\creflabelformat{equation}{(#2#1#3)}
\crefrangeformat{equation}{(#3#1#4--#5#2#6)}
\usepackage{amssymb}
\newcommand\Tau{\mathcal{T}}
\usepackage{chngcntr}
\usepackage{float}

\usepackage{lipsum} 
\usepackage{subcaption}
\usepackage{placeins} 
\usepackage{tikz}  
\usepackage{array} 

\usepackage{xcolor}
\usepackage{pifont} 
\newcommand{\cmark}{\ding{51}} 
\newcommand{\xmark}{\ding{55}} 

\usepackage{multirow}  
\usepackage{array}     
\usepackage{colortbl}  

\usepackage{titlesec}
\usepackage{lipsum} 

\begin{document}
	\title{\CH{Topological Phases of Tight-Binding Trimer Lattice in the BDI Symmetry Class}}
        \author{Mohammad Ghuneim}
        \email{g202114330@kfupm.edu.sa}
        \affiliation{%
		Department of Physics, King Fahd University of Petroleum and Minerals, 31261 Dhahran, Saudi Arabia
	}
	\author{Raditya Weda Bomantara}
	\email{Corresponding Author: Raditya.Bomantara@kfupm.edu.sa}
	\affiliation{%
		Department of Physics, Interdisciplinary Research Center for Intelligent Secure Systems, King Fahd University of Petroleum and Minerals, 31261 Dhahran, Saudi Arabia
	}
	\date{\today}
	
	
	\vspace{2cm}
	
\begin{abstract}
In this work, we theoretically study a modified Su-Schrieffer-Heeger (SSH) model in which each unit cell consists of three sites. Unlike existing extensions of the SSH model which are made by enlarging the periodicity of the (nearest-neighbor) hopping amplitudes, our modification is obtained by replacing the Pauli matrices in the system's Hamiltonian by their higher dimensional counterparts. This, in turn, leads to the presence of next-nearest neighbor hopping terms and the emergence of different symmetries than those of other extended SSH models. Moreover, the system supports a number of edge states that are protected by a combination of particle-hole, time-reversal, and chiral symmetry. Finally, our system could be potentially realized in various experimental platforms including superconducting circuits as well as acoustic/optical waveguide arrays.
	
\end{abstract}

\maketitle

\section{Introduction} 
\label{intro}
Following the revolutionary discovery of topological quantum phases \cite{Thouless82, Wen90}, the study of topological insulators has grown into a focal point within the realm of condensed matter physics \cite{Hasan2010}. Topological insulators are peculiar types of materials that display an insulating bulk, as expected from insulating materials, but their edges are conducting. Due to these materials' unique properties and potential applications, for example, in the field of quantum computing and spintronics \cite{He2022, He2019, Fan2016}, topological insulators are still
actively studied in the last few years \cite{Yang2023, Cai2023, Zhou2022, Denner2021, Tokura2019}. 
        
The Su-Schrieffer-Heeger (SSH) model \cite{Su1980} is the simplest and most fundamental model of topological insulators \cite{Asboth2016}. The SSH model describes a one-dimensional (1D) chain of unit cells, each of which contains two sites (dimers), and the coupling between intracell and intercell sites alternates. Two topologically distinct phases can be observed in the SSH structure. One of these phases, which is termed topologically nontrivial, is characterized by the presence of topologically protected zero-energy modes at the system's two edges. Such edge modes are absent in the other phase, which is thus termed topologically trivial \cite{Asboth2016}. The SSH model has been thoroughly studied both experimentally and theoretically in various physical platforms, including photonics and optical systems \cite{On2024, Liang2023, Yu2022, Roberts2022, Henriques2022, Xu2022, Aravena2022}, thermodynamic systems \cite{Upadhyay2024, Hao2022}, plasmonic systems \cite{Xia2023, Smith2021, Guan2021, Pocock2018, Downing2017}, ultracold atoms and gases \cite{Dong2024, LZhou2022, Jiang2022, Cooper2019, He2018}, ferromagnetic systems \cite{P2022, YLi2021}, acoustic waveguide \cite{Guo2024, XYang2023, QLi2023, Coutant2022, Peng2018}, and superconducting systems \cite{Banerjee2023, Rosenberg2022, Wang2021, Rosenberg2021} and circuits \cite{Zhao2023, Zheng2022, Chen2022}. Due to its simplicity, the SSH model has further been the subject for the study of topological entanglement \cite{Navarro-Labastida2022, Fromholz2020, Micallo2020}. Finally, a creative use of the SSH model that illustrates quantum state transfer is revealed in Refs.~\cite{Zurita2023, Chang2023, Zheng2023, CWang2022, Palaiodimopoulos2021, D'Angelis2020, Tan2020, Mei2018}. 

In recent years, the SSH model has undergone a transformative extension, marked by the inclusion of long-range hopping parameters \cite{Cinnirella2024, Bera2023, Dias2022, Qi2021, Pérez-González2019, Zhang2017, Malakar2023}, interaction effect \cite{Koor2022, Feng2022, Yu2020, Melo2023}, nonlinearity \cite{Jezequel2022, Y.Ma2021, Tuloup2020}, non-Hermiticity \cite{Jangjan2024, Yao2018, Lee2016, Lieu2018, Wu2021, Halder2023}, and/or periodic driving \cite{Asboth2014, Jangjan2022, Bomantara2019, Qiao2023, Wu2020, Pan2020}. Other extensions of the SSH model include the modifications of its unit cell to being trimer \cite{Verma2024, Anastasiadis2022, Alvarez2019, Du2024} or tetratomic \cite{Zhou2023, Marques2020, Xie2019}. Instead of adhering to conventional constraints of being the simplest topological insulator, such extended SSH models recognise the complexity of topological quantum phases and aim to uncover richer topological effects. This change in emphasis highlights the flexibility of the SSH model and its ability to decipher the intricacies present in quantum processes across a wide range of physical systems. This in turn puts the family of extended SSH models at the forefront of theoretical and experimental investigation of exotic topological phases. 

Our study investigates an extension of the SSH model by means of enlarging the unit cell to contain three sites (trimers). While a similar theme has been explored in previous studies such as Refs.~\cite{Verma2024, Anastasiadis2022, Alvarez2019, Du2024}, we consider a totally different approach in developing our extended SSH model. That is, instead of considering three different nearest-neighbor hopping strengths, our model is obtained by replacing the Pauli matrices in the system's Hamiltonian with their three-dimensional generalizations. As a result, our model also naturally incorporates some next-neighbor hopping terms. Due to this fundamental difference, our extended SSH model has different symmetries and topological properties from those of other trimer SSH models \cite{Verma2024, Anastasiadis2022, Alvarez2019, Du2024}. Of particular significance is the fact that our extended SSH model is protected by the chiral, particle-hole, and time-reversal symmetries, placing it in the same symmetry class (BDI) as the regular SSH model. By contrast, other existing extended SSH models typically lack the chiral symmetry and instead obey the so-called point chiral symmetry \cite{Anastasiadis2022}. \CH{Unlike chiral symmetry which pairs energy eigenvalues from the same quasimomentum sector, the point chiral symmetry pairs two opposite energy eigenvalues from different quasimomentum sectors. Intuitively, point chiral symmetry could also be understood as the manifestation of chiral symmetry acting on the $N$-dimensional bulk of a finite chain \cite{Anastasiadis2022}.} Despite having the same symmetries as the regular SSH model, the extended SSH model we propose here features richer edge state structures. Remarkably, these edge states persist despite the presence of symmetry-preserving perturbations, thus highlighting their topological nature. Finally, it is expected that our model could be experimentally implemented in various platforms that have been previously utilized to realize the SSH model. These include superconducting circuits \cite{Deng2022, Youssefi2022, Cai2019} and acoustic/optical waveguide arrays \cite{Cheng2019, Coutant2021, Yang2024}.

This paper is organized as follows. In Section~\ref{model}, we provide a detailed description of our model. There, we start by presenting the system's Hamiltonian and elucidate its physical meaning. We then perform the momentum space analysis of the system to obtain the bulk band structure and uncover its symmetries in~\ref{momentum}. In the same section, we also identify and calculate an appropriate topological invariant that dictates the system's topology. We end the section with~\ref{edge} by analytically identifying the edge states at some special parameter values, then further supporting the findings by presenting the numerically obtained edge states at more general parameter values. In Sec.~\ref{expt}, we briefly discuss the potential for realizing our system in experiments. We then consider four generic types of perturbations and investigate how they affect our model in Sec.~\ref{pert}. \CH{In section~\ref{dis}, we further demonstrate the robustness of edge states of our system in the presence of disorder.} Finally, we summarize our results and present potential aspects for future studies in Sec.~\ref{conc}.

\section{Model description}
\label{model}
\begin{center}
\begin{figure}
  \includegraphics[width=0.48\textwidth]{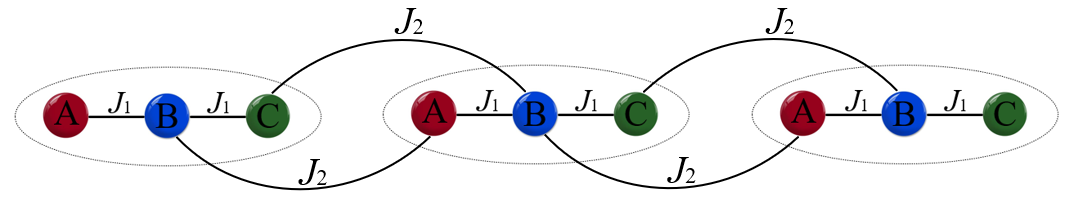}
  \captionof{figure}{Schematic of our extended lattice model with $N=3$ unit cells, A, B, and C are the three sublattices found in each unit cell.}
  \label{fig:fig1}
\end{figure}    
\end{center}

We consider a \CH{noninteracting and spinless} extended SSH model that has three sites per unit cell but only two coupling parameters; the coupling parameter $J_{1}$ controls the nearest-neighbor intracell hopping, while $J_{2}$ couples two neighboring unit cells (see Fig.~\ref{fig:fig1}). Mathematically, it is described by the Hamiltonian
\begin{flalign}
\scalebox{1.2}{$\mathscr{H}$} &= \sum_{j=1}^{N}\, \left( J_{1}\,c_{A,j}^\dagger c_{B,j} + J_{1}\,c_{B,j}^\dagger c_{C,j} + \text{\textit{h.c.}}\right) \nonumber \\
&\quad + \sum_{j=1}^{N-1}\, \left( J_{2}\,c_{A,j+1}^\dagger c_{B,j} + J_{2}\,c_{B,j+1}^\dagger c_{C,j}  + \text{\textit{h.c.}} \right), \label{eq:eq1}
\end{flalign}
where $c_{\zeta,j}^\dagger$ and $c_{\zeta,j}$ are the creation and annihilation operators at sublattice $\zeta=A,B,C$ of the $j^\text{th}$ unit cell, $N$ is the number of unit cells, $J_{1}$ and $J_{2}$ are the intracell and intercell  hopping parameters, respectively. Note that dimensionless units are used throughout this manuscript.

\subsection{Momentum space analysis}
\label{momentum}

The origin of our model construction could be better understood by writing Eq.~(\ref{eq:eq1}) in the momentum space (assuming periodic boundary conditions), i.e.,
\begin{eqnarray}
\scalebox{1.2}{$\mathscr{H}$}(k) &=& \sum_k \psi_k^\dagger h(k) \psi_k , \nonumber \\
h(k)&=&(J_{1}+J_{2}\cos(k))\,S _{x} + (J_{2}\sin(k))\, S_{y}, \label{eq:eq2}
\end{eqnarray}
where $\psi_k=(c_{A,k},c_{B,k},c_{C,k})^T$, $S_x = \left(\begin{smallmatrix} 
0 & 1 & 0\\ 
1 & 0 & 1 \\ 
0 & 1 & 0 \end{smallmatrix}\right)$ 
and 
$S_y = \left(\begin{smallmatrix} 
0 & -i & 0 \\ 
i & 0 & -i\\ 
0 & i & 0 \end{smallmatrix}\right)$ are the higher dimension generalizations of Pauli matrices. It is worth noting that $h(k)$ takes the same form as the momentum space Hamiltonian of the regular SSH model but with the Pauli matrices $\sigma_x$ and $\sigma_y$ replaced by their three-dimensional counterparts. Our construction could then be, in principle, further generalized by replacing $\sigma_x$ and $\sigma_y$ by their $D$-dimensional counterparts. However, we choose to focus on the $D=3$ case in this paper as the obtained model already exhibits rich topological effects. 

Before presenting its spectral and topological features, we shall first identify the system's symmetries. Specifically, we find that the system respects the chiral, time-reversal, particle-hole, and inversion symmetries. In terms of the momentum space Hamiltonian $h(k)$, there exists operators $\Gamma$, $\Tau$, $\mathcal{P}$, and $\mathcal{S}$ which respectively satisfy $\Gamma^{-1} h(k) \Gamma = - h(k)$, $\Tau^{-1} h(k) \Tau = h(-k)$, $\mathcal{P}^{-1} h(k) \mathcal{P} = - h(-k)$, and $\mathcal{S}^{-1} h(k) \mathcal{S} = h(-k)$. These operators are explicitly given by $\Gamma=\text{diag}(-1, 1, -1)$, $\Tau=\mathcal{K}$ ($\mathcal{K}$ being the complex conjugation operator), $\mathcal{P}=\Gamma \mathcal{K}$, and $\mathcal{S}=\begin{bmatrix} 
0 & 0 & 1 \\
0 & 1 & 0 \\
1 & 0 & 0 \\
\end{bmatrix}$. \CH{Note that $\mathcal{P}$ and $\Tau$ are anti-unitary due to the presence of the complex conjugation operator, whereas the remaining symmetries ($\Gamma$ and $\mathcal{S}$) are unitary. Moreover, observe that the chiral symmetry is related to the particle-hole and time-reversal symmetries via the relation $\Gamma=\Tau \mathcal{P}$}. It is worth emphasizing that the above symmetries are different from those respected by other trimer SSH models previously studied in Refs.~\cite{Verma2024,Anastasiadis2022, Alvarez2019, Du2024}. 
In particular, the models of Refs.~\cite{Verma2024, Anastasiadis2022, Alvarez2019} do not have chiral symmetry, whereas the model of Ref.~\cite{Du2024} lacks the particle-hole symmetry.

\begin{center}
    \begin{figure}
    \begin{subfigure}[l]{0.17\textwidth}
        \includegraphics[width=\linewidth]{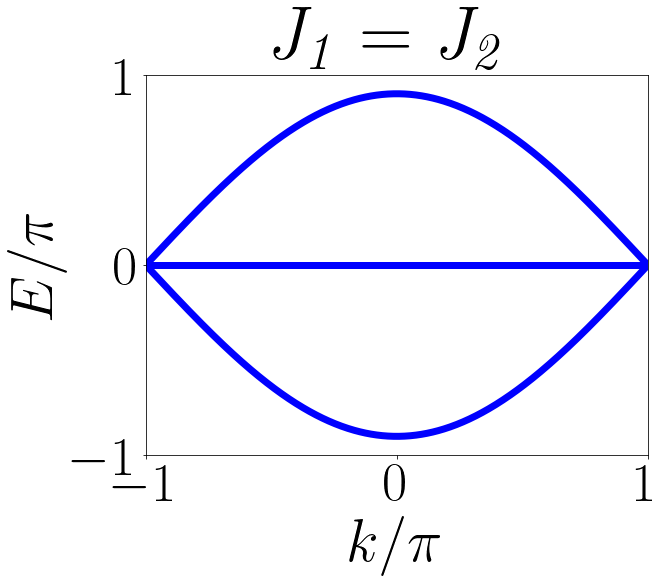}
        \caption{\texttt{}}
    \end{subfigure}%
    \hfill
    \begin{subfigure}[c]{0.155\textwidth}
        \includegraphics[width=\linewidth]{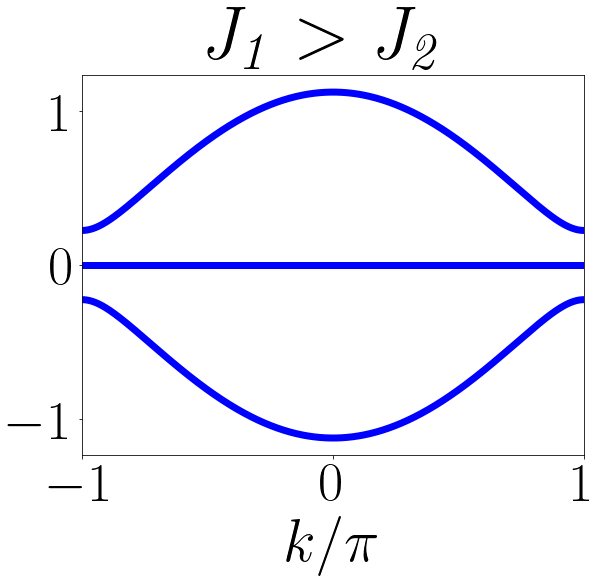}
        \caption{\texttt{}}
    \end{subfigure}%
    \hfill
    \begin{subfigure}[r]{0.155\textwidth}
        \includegraphics[width=\linewidth]{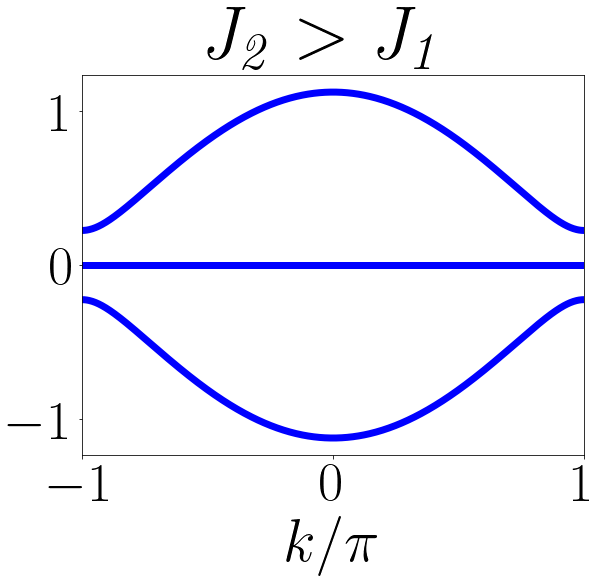}
        \caption{\texttt{}}
    \end{subfigure}
    \caption{The energy spectrum corresponding to Eq.~(\ref{eq:eq2}). (a) The energy bands touch one another at $E=0$ and $k=\pm \pi$ when $J_{1}=J_{2}$. (b,c) The bands are gapped when $J_{1}\neq J_{2}$.}
    \label{fig:fig2}
\end{figure}
\end{center}

Due to the presence of chiral symmetry $\Gamma$, the energy spectrum of our model is symmetrical about $E=0$. Indeed, the eigenvalues of $h(k)$ are calculated as follows,
\begin{eqnarray}
 E_{1}&=& 0 , \nonumber \\
 E_{2} &=& \sqrt{2(J_{1}^{2} + J_{2}^{2} + 2 J_{1} J_{2} \cos(k))} , \nonumber \\
 E_{3} &=& -\sqrt{2(J_{1}^{2} + J_{2}^{2} + 2 J_{1} J_{2} \cos(k))} .
\end{eqnarray}
These three bands are generically gapped (see Fig.~\ref{fig:fig2}), except at $J_1=J_2$ where they touch one another at the boundaries of the first Brillouin zone ($k=\pm \pi$). While the cases $J_{1}> J_{2}$ and $J_{2}> J_{1}$ appear to yield qualitatively similar spectral profiles, we will show in the following that they are topologically distinct. Moreover, our analysis in the next section further demonstrates the presence of edge states in the topologically nontrivial regime.

To unravel the topological origin of our model, we compute the normalized sublattice Zak’s phase, which was first introduced in Ref.~\cite{Anastasiadis2022} 

\begin{equation}
Z^{\zeta}_{\lambda} = \frac{i}{2}\oint dk \langle \widetilde{\psi}^{\zeta}_{\lambda}(k) | \partial_{k} \widetilde{\psi}^{\zeta}_{\lambda}(k) \rangle, \label{NSZP}
\end{equation}
where
\[
|\widetilde{\psi}^{\zeta}_{\lambda}(k) \rangle = \frac{P_{\zeta} | \psi_{\lambda}(k) \rangle}{\sqrt{\langle \psi_{\lambda}(k)| P_{\zeta} | \psi_{\lambda}(k) \rangle}},
\]
$| \psi_{\lambda}(k) \rangle$ is the eigenstate coresponding to $E_\lambda$, $\lambda=1,2,3$, and $P_{\zeta} \equiv | \zeta \rangle \langle \zeta|$ is the projector to sublattice $\zeta$, $|\zeta\rangle$ being an eigenstate of the chiral symmetry $\Gamma$ associated with the sublattice $\zeta$. \CH{Physically, as the name suggests, the normalized sublattice Zak's phase represents the Zak's phase with respect to an energy eigenstate that has been projected onto a specific sublattice (as defined by the chiral symmetry operator) and was subsequently renormalized. Such a quantity is especially significant in some chiral symmetric systems that otherwise have zero Zak's phase due to the contributions from different sublattices exactly cancelling one another. In this case, the normalized sublattice Zak's phase captures each of these contributions separately, which is further related to specific edge states in the system (depending on which energy eigenstate and sublattice projection are used in Eq.~(\ref{NSZP})). The origin of the normalized sublattice Zak's phase construction from imposing the boundary conditions on some finite chain system, as well as its explicit connection to the system's corresponding edge states under open boundary conditions, can be found in Ref.~\cite{Anastasiadis2022}.}

As detailed in Appendix~\ref{app:0}, we find that 
\[ Z^{B}_{s} =
\begin{cases}
    \pi, & \text{if } J_2 > J_1 \\
    0,   & \text{if } J_2 < J_1
\end{cases}
\]
for $s=2,3$, $Z^{B}_{1} = 0$, $Z^{A}_{\lambda}=2\,\pi$ (which is equivalent to $0$), and $Z^{C}_{\lambda}=0$. The trivial normalized sublattice Zak's phase value for the different combinations of $\zeta$ and $\lambda$ values reflects the edge states' structures of the system under open boundary conditions (OBC), which will be explicitly presented in the next section. In particular, the fact that $Z^{\zeta}_{1}=0$ for any $\zeta=A, B, C$ implies that there is no topological edge state originating from the zero energy band. On the other hand, the fact that $Z^{\zeta}_{s}=0$ for $s=2,3$, $\zeta\neq B$, and $J_2>J_1$ implies the existence of two edge states localized at a given endpoint, both of which have support on the $\zeta$ sublattice that is equal in magnitude but opposite in sign. These features will be verified in the next section.

\begin{figure}
\centering
  \includegraphics[width=0.35\textwidth]{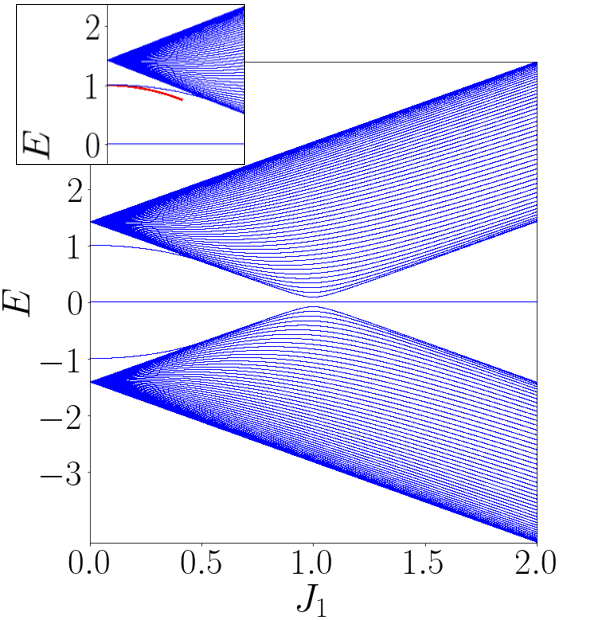}
  \captionof{figure}{ The energy spectrum of Eq.~(\ref{eq:eq1}) versus the intracell hopping amplitude $J_1$ for $N = 50$ unit cells under OBC. The intercell hopping amplitude is fixed at $J_{2}=1$. The inset shows the zoomed-in version of the energy spectrum near one of the edge states. The red curve depicts the analytically obtained edge state energy under second-order perturbation theory (see Sec.~\ref{edge}).}
  \label{fig:fig3}
\end{figure}

\begin{figure*}[htpb]
    \centering
    \begin{minipage}[b]{1\textwidth}
        \centering
        \includegraphics[width=\textwidth]{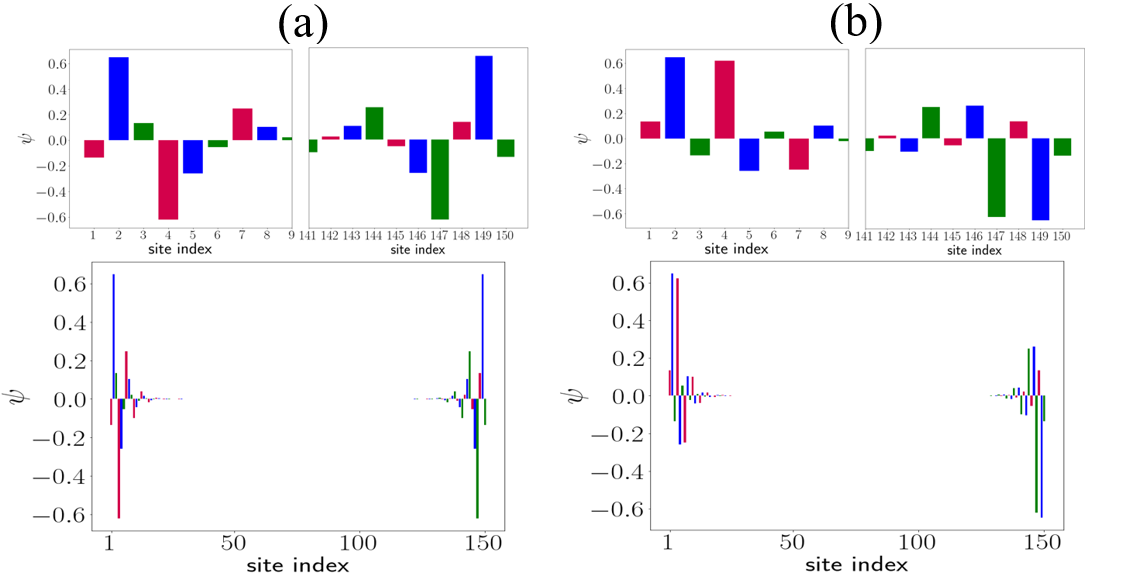}
    \end{minipage}%
   
    \caption{Edge states' profiles for Eq.~(\ref{eq:eq1}) with $J_{1}=0.2$, $J_{2}=1$, and $N=50$ unit cells. Subplot (a) shows the distinct left and right edge states at $E\approx-0.96$ and subplot (b) illustrates the two edge states at $E\approx 0.96$. The three sublattices are marked with different colors according to the schematic of Fig.~\ref{fig:fig1}. The upper panels show the zoomed-in view of each edge state near the appropriate (left or right) lattice edge.}
    \label{fig:fig4}
\end{figure*}

\subsection{Formation of edge states in the real space}
\label{edge}

Having identified the band closing location, i.e., at $J_1=J_2$, that separates the two distinct topological phases in the previous section, we now turn our attention again to the real space description of Eq.~(\ref{eq:eq1}). In particular, we aim to demonstrate the correlation between topology and the presence of edge states, i.e., the topologically nontrivial regime $J_2>J_1$ supports robust localized eigenstates near the system's edges, whilst such edge states are absent in the topologically trivial regime $J_1>J_2$.

We start by writing Eq.~(\ref{eq:eq1}) under OBC in the form
\begin{equation}
    \mathcal{H} = \left(\begin{array}{ccccc}
         c_{A,1}^\dagger & c_{B,1}^\dagger & c_{C,1}^\dagger & \cdots & c_{C,N}^\dagger \\
    \end{array}\right) \mathcal{H} \left(\begin{array}{c}
         c_{A,1}  \\
         c_{B,1}  \\         
         c_{C,1}  \\
         \vdots \\
         c_{C,N}
    \end{array}\right),
    \label{eq:eq5}
\end{equation}
where $\mathcal{H}$ is a $3N\times 3N$ matrix of the form

\begin{flalign}
 \mathcal{H} &= \sum_{j=1}^{N}\, \left( J_{1}\,| A,j \rangle \langle B,j | + J_{1}\,| B,j \rangle \langle C,j | + \text{\textit{h.c.}}\right) \nonumber \\ 
 &\quad + \sum_{j=1}^{N-1}\, \left( J_{2}\,| A,j+1 \rangle \langle B,j | + J_{2}\,| B,j+1 \rangle \langle C,j |  + \text{\textit{h.c.}} \right) , 
\label{eq:eq6}
\end{flalign}
$|S,j\rangle $ is a $3N\times 1$ column vector whose elements are 1 at the $(3j-s)$th row ($s=2,1,0$ for $S=A,B,C$) and 0 elsewhere. By numerically diagonalizing $\mathcal{H}$ above, we obtain the system's energy spectrum in Fig.~\ref{fig:fig3}. There, we verify that the bands are gapped, except at $J_1=J_2$. Moreover, in-gap states are clearly observed in the regime $J_1<J_2$, as we previously claimed. In Fig.~\ref{fig:fig4}, the wave function profiles of such in-gap states at fixed $J_1$ and $J_2$ values demonstrate their localized nature near the left or right end. This shows that, despite not being pinned at $E=0$ as in the regular SSH model, the observed in-gap states are in fact the sought-after edge states. Upon closer inspection of Fig.~\ref{fig:fig3}, the energy of the observed edge states is quadratic with respect to $J_1$ and is exactly $E=\pm J_2$ at $J_1=0$. This insight allows us to support the above numerical results with the following analytical treatment.

To obtain the system's edge states analytically, we start by setting $J_1=0$ and solving the eigenvalue equation $\mathcal{H} \ket{\psi} = E \ket{\psi}$ with $E=\pm J_2$. By deferring the mathematical details to Appendix~\ref{app:A}, we obtain (ignoring any normalization constant):
\[
\ket{\psi_{L,\pm}} =
|B,1 \rangle \pm |A,2 \rangle ,
\]
for the left localized edge states, and
\[
\ket{\psi_{R,\pm}} =
|B,N \rangle \pm |C,N-1 \rangle , 
\]
for the right localized ones.

At nonzero values of $J_1$, the energy of the edge states could be estimated via perturbation theory. In particular, by taking the intracell hopping term as the perturbative potential, i.e., $\mathcal{H}_{\text{o}} \equiv \sum_{j=1}^{N}\, \left( J_{1}\,| A,j \rangle \langle B,j | + J_{1}\,| B,j \rangle \langle C,j | + \text{\textit{h.c.}}\right)$, we find that the first-order energy correction for both edge states is $\Delta E_{L,\pm}^{(1)} = \langle \psi_{L,\pm} | \mathcal{H}_{\text{o}} | \psi_{L,\pm} \rangle =0$. To obtain the lowest nonzero correction, we then evaluate the second-order correction as
\begin{eqnarray}
\Delta E_{L,\pm}^{(2)} &=& \sum_{m\neq (L,\pm)} \frac{\left| \langle \psi_{L,\pm} | \mathcal{H}_{\text{o}} | \psi_{m} \rangle \right|^2}{E_{m}^{(0)} - E_{L,\pm}^{(0)}} \nonumber \\
&\approx & \frac{\left| \langle \psi_{L,\pm} | \mathcal{H}_{\text{o}} | \psi_{L,\mp} \rangle \right|^2}{E_{L,\mp}^{(0)} - E_{L,\pm}^{(0)}} = \mp\,2 \frac{ | J_{1}|^2}{J_{2}}.
\label{eq:eq7}
\end{eqnarray}
 In the second equality above, we ignored all but one term in the summation. Specifically, we considered only the term coming from the other edge state localized in the same edge since it contains the largest overlap $\left| \langle \psi_{L,\pm} | \mathcal{H}_{\text{o}} | \psi_{L,\mp} \rangle \right|^2$. Indeed, as demonstrated in the inset of Fig.~\ref{fig:fig3}, excellent agreement is observed between the analytically obtained edge states under second-order perturbative approximation above and the numerically obtained ones.
 
\section{Discussion}

\subsection{Potential experimental realizations}
\label{expt}

Due to its simplicity, the regular SSH model has been realized in various experimental platforms, including superconducting circuits \cite{Deng2022, Youssefi2022, Cai2019} and acoustic/optical waveguide arrays \cite{Sougleridis2024, Cheng2019, Coutant2021, Yang2024}. It is expected that, with suitable modifications, these experiments could be adapted to realize our extended SSH model. 

\begin{figure}
  \centering
  \includegraphics[width=0.2\textwidth]{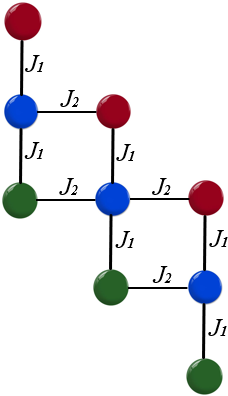}
  \captionof{figure}{The quasi-1D configuration for our extended SSH model which renders all couplings nearest-neighbor.}
  \label{fig:fig5}
\end{figure}

The potentially nontrivial component of our extended SSH model that is not present in the regular SSH model is the next-nearest neighbor coupling, which necessarily arises due to the intercell hopping. Fortunately, such a next-nearest neighbor coupling could be handled in at least two different ways. First, specific to the optical waveguide platform, a next-nearest neighbor coupling could be achieved by using waveguide interference phenomena \cite{Savelev2020}. That is, we may efficiently link waveguides at extended distances by taking advantage of interference processes involving an extra supplemental waveguide in between. Alternatively, a potentially simpler means of handling the next-nearest neighbor couplings is by turning them into nearest neighbor ones. This could be achieved by rearranging the lattice configuration into a quasi-1D ladder as illustrated in Fig.~\ref{fig:fig5}.

The extended SSH model under Fig.~\ref{fig:fig5} configuration could be implemented in existing experimental platforms. For example, in superconducting circuits, the sites could represent Xmon qubits, whereas the coupling between two neighboring sites could be achieved inductively by a tunable coupler \cite{Mei2018} or capacitively such as in \cite{Cai2019}. In acoustic/optical waveguides, each lattice site is represented by a waveguide, the propagation direction of which simulates the arrow of time. In this case, the coupling between two neighboring waveguides is achieved by a tunneling effect that can be controlled by their separation \cite{XinLi2018}. The presence of edge states in the system could then be verified by exciting appropriate waveguides near a lattice edge and tracking the propagation dynamics.

\subsection{Effects of perturbation}
\label{pert}

Topological edge states are particularly attractive for their robustness against perturbations that preserve the system's protecting symmetries. In the following, we further support the topological signature of the system and identify its protecting symmetries by investigating the fate of its edge states in the presence of some representative perturbations. To this end, each of the following terms shall be separately added to the momentum space Hamiltonian of Eq.~(\ref{eq:eq2}),

\begin{eqnarray}
    \alpha(k) \equiv \alpha_{\text{o}} \scalebox{0.6}{$\begin{pmatrix} 0 & -i & 0 \\ i & 0 & 0 \\ 0 & 0 & 0 \end{pmatrix}$} &,& \beta(k) \equiv \beta_{\text{o}} \scalebox{0.6}{$\begin{pmatrix} 0 & e^{-ik} & 0 \\ e^{ik} & 0 & 0 \\ 0 & 0 & 0 \end{pmatrix}$}, \nonumber \\
    \gamma(k) \equiv \gamma_{\text{o}} \scalebox{0.6}{$\begin{pmatrix} 0 & 1 & 0 \\ 1 & 0 & 0 \\ 0 & 0 & 1 \end{pmatrix}$} &,& 
   \delta(k) \equiv \delta_{\text{o}} \scalebox{0.6}{$\begin{pmatrix} 1 & 0 & 0 \\ 0 & -1 & 0 \\ 0 & 0 & 1 \end{pmatrix}$} ,
   \label{eq:eq8}
\end{eqnarray}
where $\alpha_{\text{o}}$, $\beta_{\text{o}}$, $\gamma_{\text{o}}$, and $\delta_{\text{o}}$ are the respective perturbation strengths. Under each of these perturbations, the respective real space Hamiltonian reads,

\begin{flalign}
\scalebox{1.2}{$\mathscr{H}_\alpha$} &= \sum_{j=1}^{N}\, \left[(J_{1}-i\alpha_{\text{o}})\,c_{A,j}^\dagger c_{B,j} + J_{1}\,c_{B,j}^\dagger c_{C,j} \right] \nonumber \\
&\quad  + \sum_{j=1}^{N-1} \left[ J_{2}\,c_{A,j+1}^\dagger c_{B,j} + J_{2}\,c_{B,j+1}^\dagger c_{C,j} \right] + \text{\textit{h.c.}} ,
\label{eq:eq9}
\end{flalign}

\begin{flalign}
\scalebox{1.2}{$\mathscr{H}_\beta$} &= \sum_{j=1}^{N}\, \left[ J_{1}\,c_{A,j}^\dagger c_{B,j} + J_{1}\,c_{B,j}^\dagger c_{C,j} \right] \nonumber \\ &\quad + \sum_{j=1}^{N-1} \left[ (J_{2}+\beta_{\text{o}})\,c_{A,j+1}^\dagger c_{B,j} + J_{2}\,c_{B,j+1}^\dagger c_{C,j} \right]  + \text{\textit{h.c.}} ,
\label{eq:eq10}
\end{flalign}

\begin{flalign}
\scalebox{1.2}{$\mathscr{H}_\gamma$} &= \sum_{j=1}^{N}\, \left[ (J_{1}+\gamma_{\text{o}}) c_{A,j}^\dagger c_{B,j} + J_{1}\,c_{B,j}^\dagger c_{C,j} \right] \nonumber \\
&\quad + \sum_{j=1}^{N-1} \left[ J_{2}\,c_{A,j+1}^\dagger c_{B,j} + J_{2}\,c_{B,j+1}^\dagger c_{C,j} \right]  + \text{\textit{h.c.}} \nonumber \\ &\quad + \gamma_{\text{o}}\sum_{j=1}^{N}\,c_{C,j}^\dagger c_{C,j} , 
\label{eq:eq11}
\end{flalign}

\begin{flalign}
\scalebox{1.2}{$\mathscr{H}_\delta$} &= \sum_{j=1}^{N}\, \left[ J_{1}\,c_{A,j}^\dagger c_{B,j} + J_{1}\,c_{B,j}^\dagger c_{C,j} \right] \nonumber \\ &\quad + \sum_{j=1}^{N-1} \left[J_{2}\,c_{A,j+1}^\dagger c_{B,j} + J_{2}\,c_{B,j+1}^\dagger c_{C,j} \right] + \text{\textit{h.c.}} \nonumber \\
&\quad +\delta_{\text{o}}\sum_{j=1}^{N} \, \left[ c_{A,j}^\dagger \,c_{A,j} - \,c_{B,j}^\dagger \,c_{B,j} + \,c_{C,j}^\dagger \,c_{C,j} \right]. 
\label{eq:eq12}
\end{flalign}

The perturbations $\alpha(k)$ and $\beta(k)$ have the effect of introducing an imbalance between the two intracell and intercell hopping amplitudes respectively. \CH{In particular, the imaginary nature of $\alpha(k)$ is made to result in a complex intracell hopping. Physically, such a complex hopping represents a relative phase difference between the intracell and intercell hoppings, which could arise in the design of Fig.~\ref{fig:fig5} when the formed square loops are subject to some flux.} It is easily verified that both $\alpha(k)$ and $\beta(k)$ preserve the chiral symmetry since $\Gamma^{(-1)}\alpha(k) \Gamma =-\alpha(k)$ and $\Gamma^{(-1)}\beta(k) \Gamma =-\beta(k)$. In contrast to the perturbation $\alpha(k)$ which breaks time-reversal, particle-hole, and inversion symmetries, $\beta(k)$ preserves time-reversal and particle-hole symmetries while breaking inversion symmetry.

The perturbation $\gamma(k)$ is chosen to introduce an on-site potential on sublattice $C$, while at the same time further introducing imbalance between the two intracell hopping amplitudes. The perturbation $\delta(k)$ is chosen to introduce on-site potentials on all three sublattices. Both $\gamma(k)$ and $\delta(k)$ break the chiral and particle-hole symmetries while preserving time-reversal symmetry. On the other hand, the inversion symmetry is broken by $\gamma(k)$ and is preserved by $\delta(k)$. \CH{The effect of all the introduced perturbations on the system symmetries is summarized in Table~\ref{tbl:tbl1}.}

\begin{table}[ht]
    \centering
    \resizebox{0.35\textwidth}{!}{
    \begin{tabular}{|>{\columncolor{white}}c|c|c|c|c|}
        \hline
        \begin{tikzpicture}
            
            \draw (1.0,-1.4) -- (-2.6,0.0);
            
            \node[anchor=center] at (-1.6, -1.1) {\textbf{Symmetry}};
            \node[anchor=center] at (0.15, -0.5) {\textbf{Operator}};
        \end{tikzpicture}
        & \raisebox{5mm}{$\alpha(k)$} 
        & \raisebox{5mm}{$\beta(k)$} 
        & \raisebox{5mm}{$\gamma(k)$} 
        & \raisebox{5mm}{$\delta(k)$}    \\ 
        \hline
        Chiral & \cmark & \cmark & \xmark & \xmark \\
        \hline
        Time-reversal & \xmark & \cmark & \cmark & \cmark \\
        \hline
        Particle-hole & \xmark & \cmark & \xmark & \xmark \\
        \hline
        Inversion & \xmark & \xmark  & \xmark & \cmark \\
        \hline
    \end{tabular}}
    \caption{\CH{Summary of symmetries preserved and broken by each perturbation operator considered in Sec.~\ref{pert}. Our normalized sublattice Zak's phase calculations (see Fig.~\ref{fig:fig7b}) only correctly predict the edge states under perturbations $\beta(k)$ and $\delta(k)$.}}
    \label{tbl:tbl1}
\end{table}

\begin{figure}[htpb]
  \centering
  \includegraphics[width=0.4\textwidth]{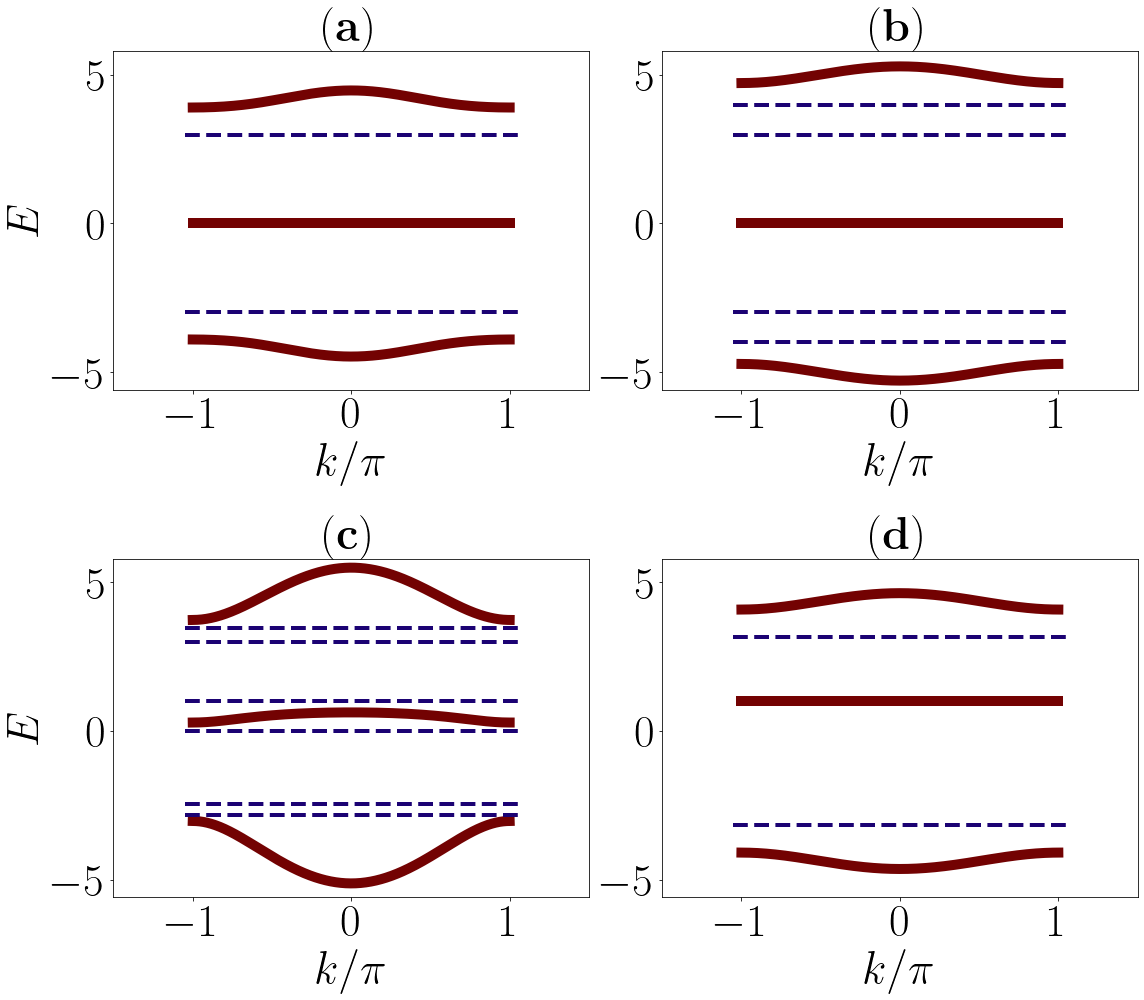}
  \captionof{figure}{The momentum space energy spectrum in the presence of the four perturbations $\alpha(k)$, $\beta(k)$, $\gamma(k)$, and $\delta(k)$. The system parameters are chosen as $J_{1}=0.2$, $J_2=3$ and (a) $\alpha_0=1$, (b) $\beta_0=1$, (c) $\gamma_0=1$, and (d) $\delta_0=1$. The blue dashed lines mark the energies of the expected edge states when OBC are applied (see Fig.~\ref{fig:fig6})} 
  \label{fig:fig6b}
\end{figure}

\begin{figure}[htpb]
  \centering
  \includegraphics[width=0.45\textwidth]{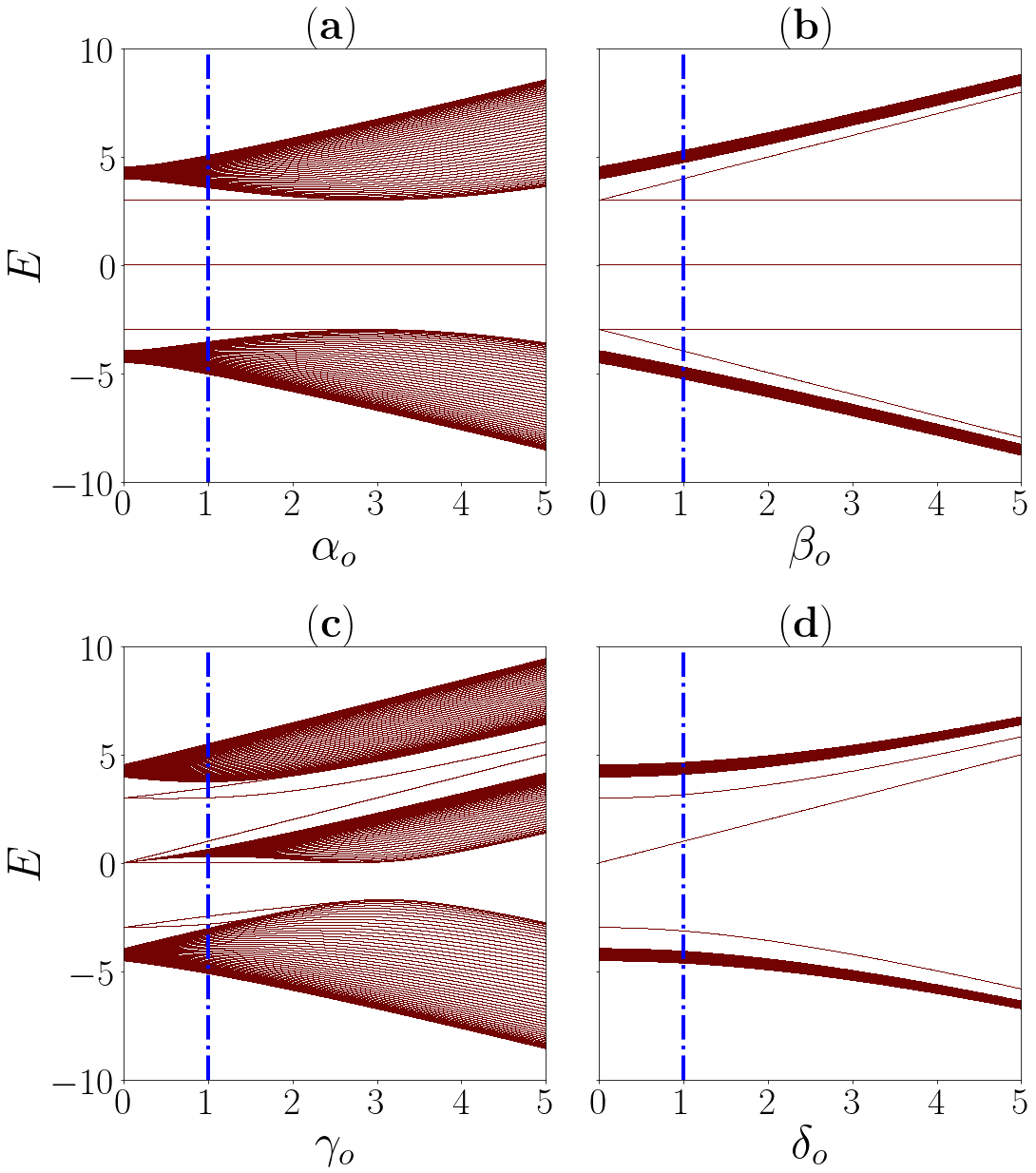}
  \captionof{figure}{The real space energy spectrum versus the perturbation strengths for Eqs.~\Crefrange{eq:eq9}{eq:eq12} in a system of $N=50$ unit cells. The other parameters are chosen as $J_{1}=0.2$ and $J_{2}=3$ for all cases. The blue vertical lines mark the parameter values used in Fig.~\ref{fig:fig6b}.}
  \label{fig:fig6}
\end{figure}

\begin{figure}
  \centering
  \includegraphics[width=0.45\textwidth]{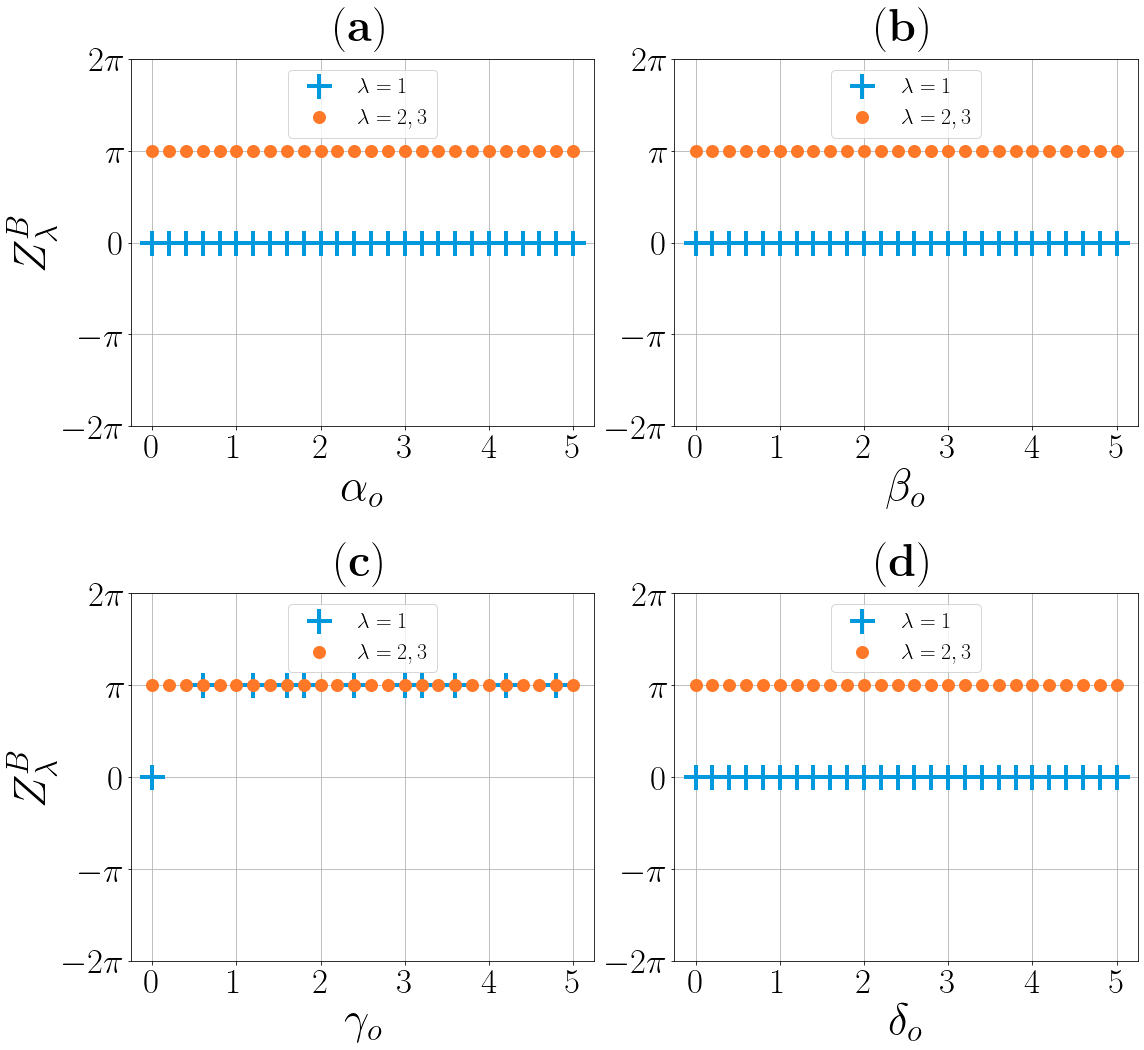}
  \captionof{figure}{The normalized sublattice Zak's phase in the presence of the four perturbations. The system parameters are $J_{1}=0.2$ and $J_{2}=3$ in all cases.}
  \label{fig:fig7b}
\end{figure}

\begin{figure}
  \centering
  \includegraphics[width=0.45\textwidth]{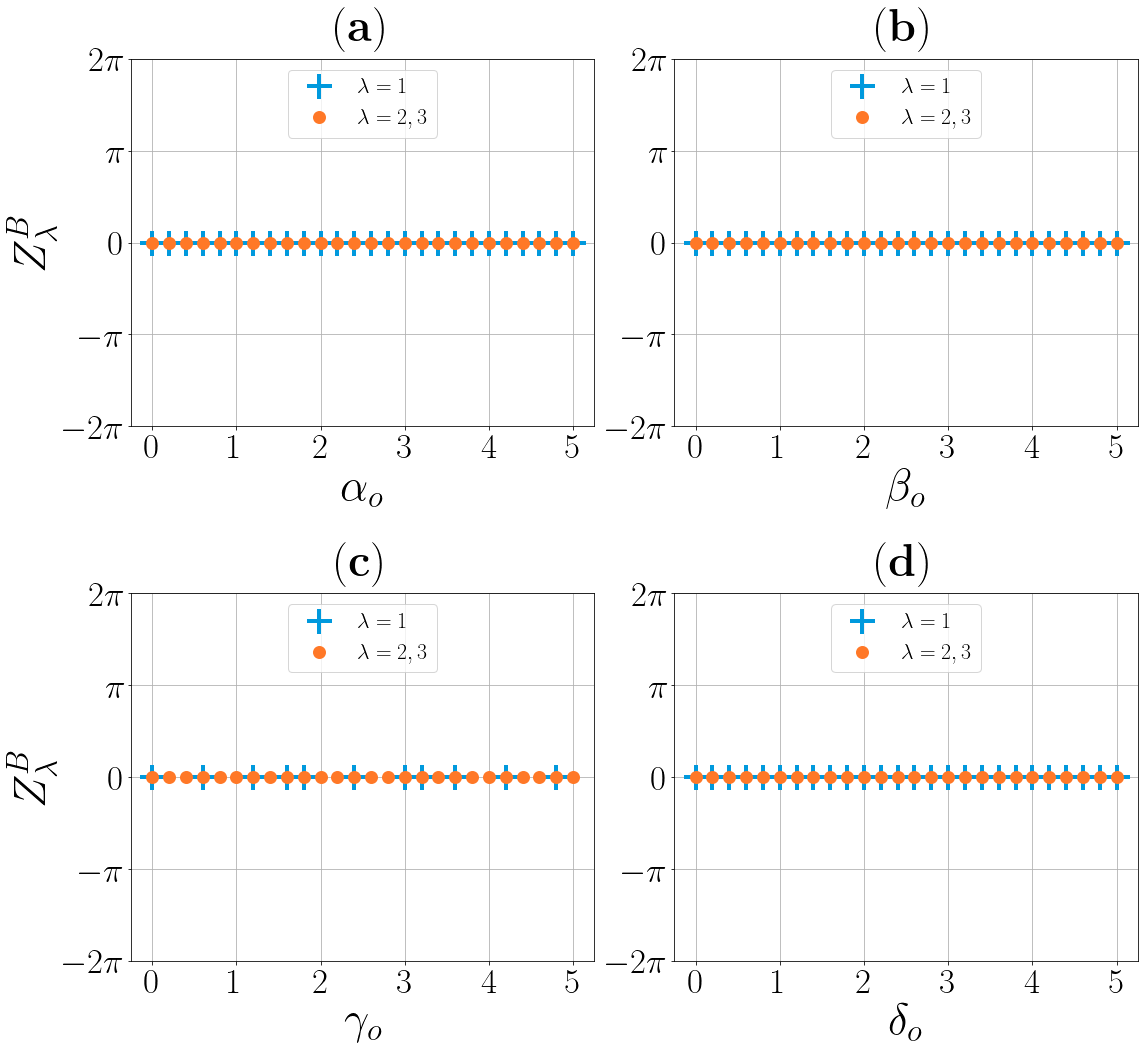}
  \captionof{figure}{\CH{The normalized sublattice Zak's phase in the presence of the four perturbations. The system parameters are $J_{1}=3$ and $J_{2}=0.2$ in all cases.}}
  \label{fig:fig7c}
\end{figure}

Figures~\ref{fig:fig6b} and \ref{fig:fig6} show the energy spectra of our trimer SSH model in the presence of each of the aforementioned perturbations in the momentum space and the real space respectively. Upon inspecting the two figures and noting that only bulk states are reflected in the momentum space energy spectrum of Fig.~\ref{fig:fig6b}, we are able to distinguish between bulk and edge states in the real space energy spectrum of Fig.~\ref{fig:fig6}. By further tracking the edge states at varying perturbation strengths, the roles of the system's four symmetries could further be uncovered. 

As shown in Fig.~\ref{fig:fig6}(a), we find that the perturbation $\alpha(k)$ may lead to the disappearance of the edge states without being accompanied by band closing. The opposite effect is observed for the perturbation $\gamma(k)$ in Fig.~\ref{fig:fig6}(c) whereby new edge states spontaneously emerge from the middle bulk band at finite perturbation strength. On the other hand, when chiral, particle-hole, and time-reversal symmetries are preserved, edge states do not spontaneously appear or disappear, as is the case in Fig.~\ref{fig:fig6}(b). These results demonstrate that the simultaneous presence of chiral, particle-hole, and time-reversal symmetries is necessary for the full topological protection of the system, thus placing it in the BDI class of the Altland-Zirnbauer classification \cite{AZ97}. However, it is worth noting that breaking some of these symmetries does not always lead to the spontaneous destruction or emergence of the edge states (see, e.g., Fig.~\ref{fig:fig6}(d)).

While the perturbation $\beta(k)$ preserves the system's edge states, it breaks the degeneracy between the left and right edge states (see Fig.~\ref{fig:fig6}(b)). In this case, while the energy of the right edge states remains unaffected by the perturbation strength $\beta_{\text{o}}$, the energy of the left edge states is shifted closer to the nearest bulk band. Consequently, at large enough $\beta_{\text{o}}$, the left edge states merge with the bulk bands and disappear, whilst the right edge states remain present. Such an asymmetry between the number of edge states on the left and right edge states is unique to our extended SSH model and cannot be found in a regular SSH model (without breaking the sublattice degree of freedom). This phenomenon could be understood from Eq.~(\ref{eq:eq8}) by observing that the perturbation $\beta(k)$ only modifies the hopping involving sublattices A and B from two neighboring unit cells. As the left edge states have the largest support on sublattices A and B (see Fig.~\ref{fig:fig4}), they are significantly affected by such a perturbation. By contrast, the right edges, which have the largest support on sublattices B and C (see Fig.~\ref{fig:fig4}), are much more insensitive to $\beta(k)$. 

As the perturbation $\gamma(k)$ also modifies the hopping involving sublattices A and B while leaving the hopping involving sublattices B and C intact, the left and right edge states similarly develop an energy difference. Moreover, as the chiral symmetry is broken, there is a further asymmetry between the positive and negative energies. In particular, as observed in Fig.~\ref{fig:fig6}(c), both the negative energy edge states eventually merge with the lower bulk band and disappear at large enough $\gamma_{\text{o}}$. On the other hand, at positive energy, only the right localized edge state merges with the upper bulk, whilst the left localized edge state persists even at very large $\gamma_{\text{o}}$. Interestingly, the perturbation $\gamma(k)$ also leads to seemingly new edge states emerging from the zero energy bulk states. \CH{It should be noted that these ``new" edge states actually also exist in the unperturbed case at zero energy, which therefore coexist with the bulk states. This could be easily seen by inspecting the schematics of Fig.~\ref{fig:fig1} and taking $J_1=0$, in which case the left- and right-most sites are decoupled and thus correspond to zero energy edge states. As $\gamma(k)$ introduces an onsite potential on sublattice C, the bulk states and the right-localized edge state are shifted from zero energy, thus breaking their degeneracy. As the left-localized edge state has negligible support on sublattice C, it remains at zero energy. At moderate $\gamma_{\text{o}}$, however, this edge state merges back with the center bulk band that has now gained a significant bandwidth. On the other hand, the right-localized edge state} remains present at very large $\gamma_{\text{o}}$. It then follows that at $\gamma_0 \rightarrow \infty$, only two (almost degenerate) edge states at positive energy remain, one localized near the left edge whilst the other is localized near the right edge of the system. 

Finally, despite breaking all the protecting symmetries, the perturbation $\delta(k)$ merely deforms the energy bands while leaving the edge states intact (see Fig.~\ref{fig:fig6}(d)). The breaking of the chiral symmetry again manifests itself as the asymmetry in the energy spectrum around $E=0$ that results from the center bulk band shifting upward. Nevertheless, the edge states remain well-gapped from the bulk bands even at large $\delta_{\text{o}}$, thus ruling out the possibility for these edge states to merge with the bulk bands and disappear as in Fig.~\ref{fig:fig6}(a).   

For completeness, in Appendix~\ref{app:B}, we show the wave function profiles of the various edge states under the different perturbations above. In Fig.~\ref{fig:fig7b}, we further numerically calculate the normalized sublattice Zak's phase $Z_\lambda^B$ introduced in Sec.~\ref{momentum} at varying perturbation strengths. \CH{While the computation of normalized sublattice Zak's phase only involves the chiral symmetry, its correlation with the system's edge states is only guaranteed if chiral, particle-hole, and time-reversal symmetries are \emph{simultaneously} present. Indeed, we observe that this is the case for perturbation $\beta$, i.e., the quantization of $Z_\lambda^B$ to $\pi$ for the lowest and highest bands is consistent with the persistence of the system's edge states at large perturbation strengths ($\beta_0$ and $\delta_0$).} On the other hand, $Z_\lambda^B$ is no longer able to faithfully determine the existence of edge states in the presence of perturbations $\alpha$ and $\gamma$, \CH{which break some of the protecting (chiral, particle-hole, and/or time-reversal) symmetries}.  

 \CH{Specifically, while $\alpha$ preserves chiral symmetry, it breaks time-reversal symmetry.} This explains why at large $\alpha_0$ values, $Z_\lambda^B$ falsely yields a nontrivial value even when the system no longer appears to support edge states. In the case of perturbation $\gamma$ which breaks chiral symmetry, it is further no longer guaranteed that $Z_\lambda^B$ remains quantized or even well-defined. Indeed, Fig.~\ref{fig:fig7b}(c) reveals that $Z_\lambda^B$ becomes ill-defined at some $\gamma_0$ values. Interestingly, at $\gamma_0$ values where $Z_\lambda^B$ remains well-defined, $Z_1^B$ takes on the nontrivial $\pi$ value instead of 0 as in the unperturbed case. This could be attributed to the fact that $\gamma$ perturbation results in additional edge states emerging from the middle band, as evidenced from Fig.~\ref{fig:fig6}(c). \CH{It should be noted that while normalized sublattice Zak's phase no longer accurately correlates with the system's edge states in the presence of symmetry-breaking perturbations, it does not mean that it always fails in capturing the system's edge states. Curiously, for some specific types of symmetry-breaking perturbations, e.g., under perturbation $\delta$, a good agreement could still be found between the normalized sublattice Zak's phase and the system's edge states. This could be attributed to the special form of perturbations under consideration; in our case, $\delta$ is simply the chiral symmetry operator itself. Finally, Fig.~\ref{fig:fig7c} shows the numerically computed normalized sublattice Zak's phase $Z_\lambda^B$ as the perturbation strengths vary for the case when $J_{1} > J_{2}$. As expected for the trivial regime, $Z_\lambda^B$ exhibits a trivial value at all perturbation strengths in all cases.}

\subsection{Robustness against disorder}
\label{dis}

\CH{One attractive feature of topological insulators is the robustness of their edge states against mild degrees of spatial disorder. To further support the topological nature of our system, it is thus instructive to investigate the fate of the above-obtained edge states in the presence of such disorder. To this end, we modify each hopping parameter $J_{i}$ (where $i=1,2$) by adding a random (site-dependent) component $d_{j}$ such that $J_{i,d_{j}}= J_{i} + d_{j}$, where $j=1,2,\cdots,N$ and $d_{j}$ is uniformly distributed in some [$-\Delta$, $\Delta$]. Here, $\Delta$ represents the maximum amplitude of the disorder for the hopping parameter $J_{i}$ and is referred to as the disorder strength. The random part $d_{j}$ alters the value of the hopping parameter connecting each pair of sites by increasing or decreasing it, emulating a situation in which fluctuations in the hopping strengths are brought either by flaws or the system's local design. For completeness, we will first introduce disorder on both hopping parameters simultaneously to investigate their combined impacts that are particularly relevant in actual experiments, then we investigate the case of isolated disorder in either $J_{1}$ or $J_{2}$ separately in order to better understand the effect of disorder in each hopping individually.}


\begin{figure}[h]
    \centering
    \begin{subfigure}[b]{0.23\textwidth}
        \centering
        \includegraphics[width=\textwidth]{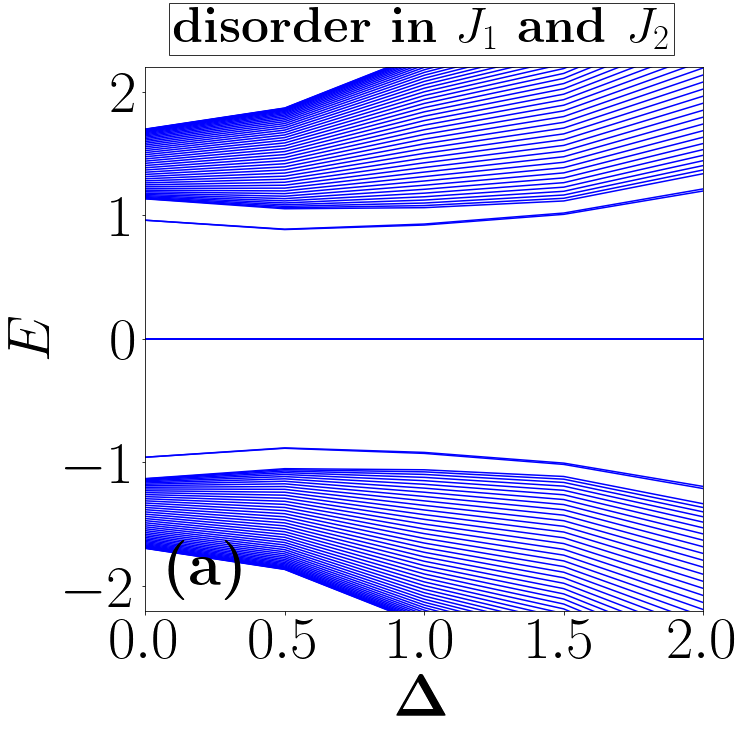}
        \label{fig:sub1}
    \end{subfigure}
    \begin{subfigure}[b]{0.23\textwidth}
        \centering
        \includegraphics[width=\textwidth]{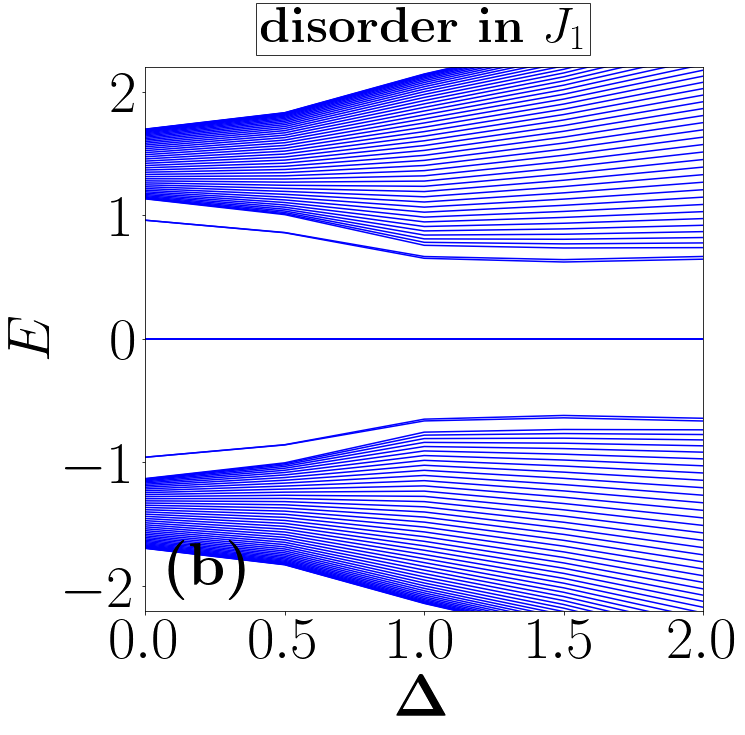}    
        \label{fig:sub2}
    \end{subfigure}
    
    \vspace{0.0cm} 
    
    \begin{subfigure}[bc]{0.23\textwidth}
        \centering
        \includegraphics[width=\textwidth]{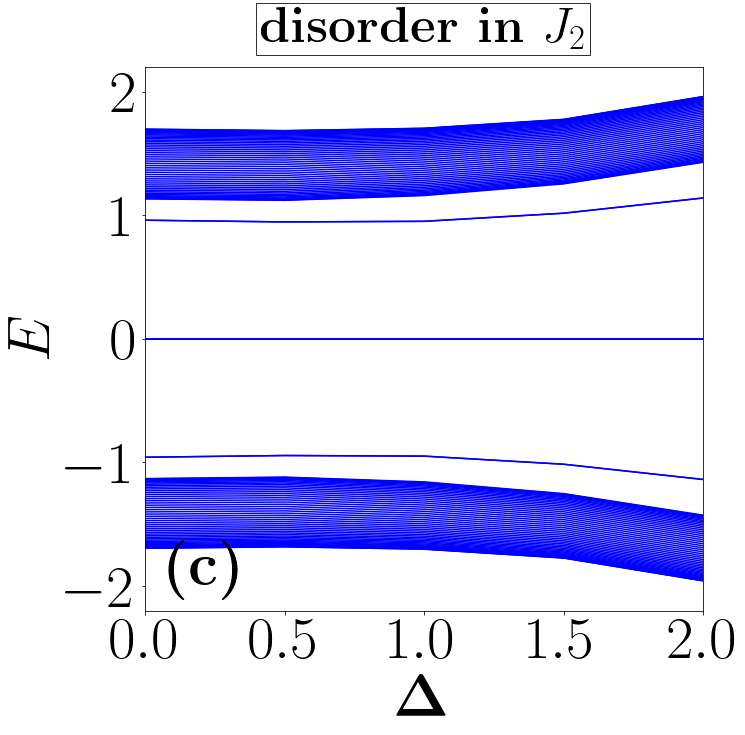}    
        \label{fig:sub3}
    \end{subfigure}
    
    \caption{\CH{The energy spectrum of disordered Eq.~(\ref{eq:eq1}) versus the disorder strength for $N = 50$ unit cells under OBC. The starting parameter values for all three cases are $J_{1} = 0.2$ and $J_{2} = 1$. (a) Disorder acts on both $J_{1}$ and $J_{2}$, (b) Disorder acts only on  $J_{1}$, and (c) Disorder acts only on $J_{2}$. Each data point is averaged over 200 disorder realizations.}}
    \label{fig:fig1d}
\end{figure}

\CH{Our results are summarized in Fig.~\ref{fig:fig1d}, which generally confirm the robustness of the system's edge states in the presence of disorder. Indeed, as shown in Fig.~\ref{fig:fig1d}(a), the edge states remain visible and well-separated from their nearest bulk band even in the presence of considerable disorder in both intracell and intercell hopping. It is also interesting to note that disorder yields different effects on intracell and intercell hopping. In particular, Fig.~\ref{fig:fig1d}(b) reveals that disorder in the intracell hopping tends to reduce the gap between the edge states and their nearest bulk band, thus reducing their visibility. By contrast, as demonstrated in Fig.~\ref{fig:fig1d}(c), the disorder in the intercell hopping enhances the visibility of the edge states by increasing the gap with their nearest bulk band. Intuitively, the different effects of disorder in the two hopping parameters could be attributed to the fact that disorder in the intracell (intercell) hopping has the tendency to effectively bring the system to the topologically trivial (nontrivial) regime of $J_1>J_2$ ($J_1<J_2$), thus explaining its negative (positive) effect on the system's edge states. }


\begin{figure}[h]
    \centering
    \begin{subfigure}[b]{0.23\textwidth}
        \centering
        \includegraphics[width=\textwidth]{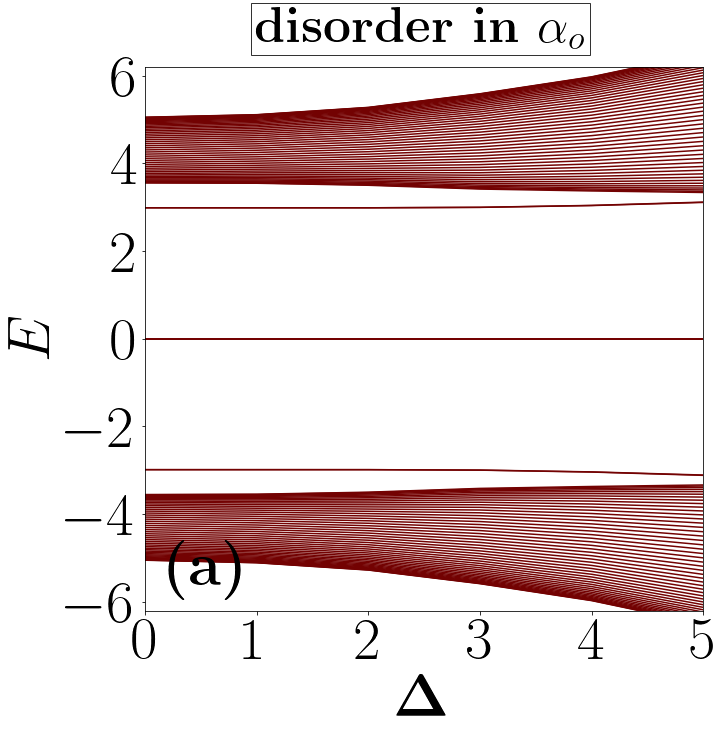}
        \label{fig:sub1}
    \end{subfigure}
    \begin{subfigure}[b]{0.23\textwidth}
        \centering
        \includegraphics[width=\textwidth]{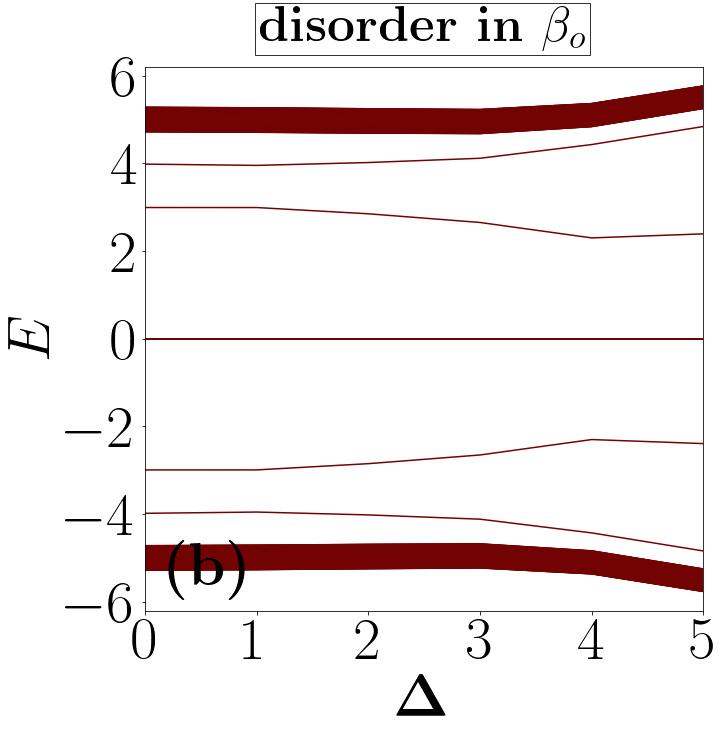}      
        \label{fig:sub2}
    \end{subfigure}   
    \begin{subfigure}[b]{0.23\textwidth}
        \centering
        \includegraphics[width=\textwidth]{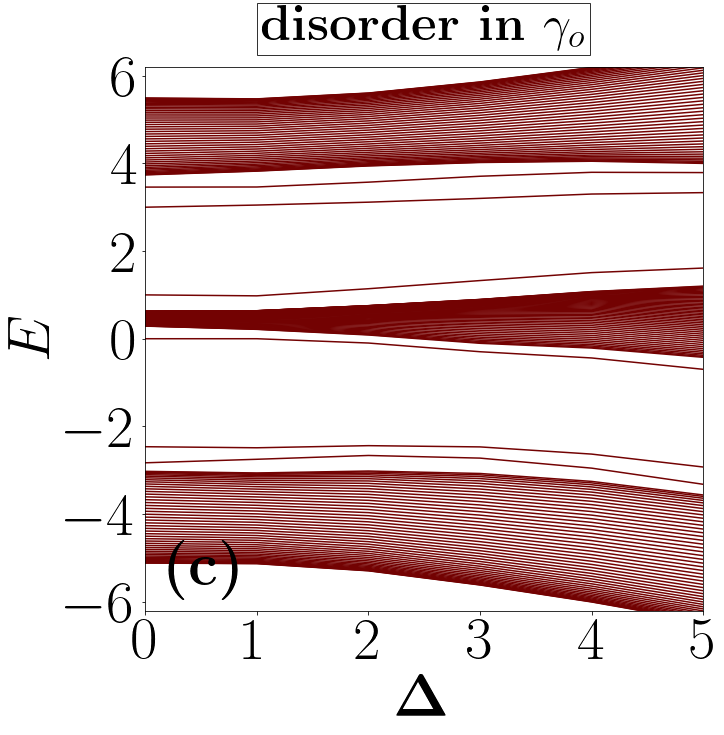}
        \label{fig:sub3}
    \end{subfigure}
    \begin{subfigure}[b]{0.23\textwidth}
        \centering
        \includegraphics[width=\textwidth]{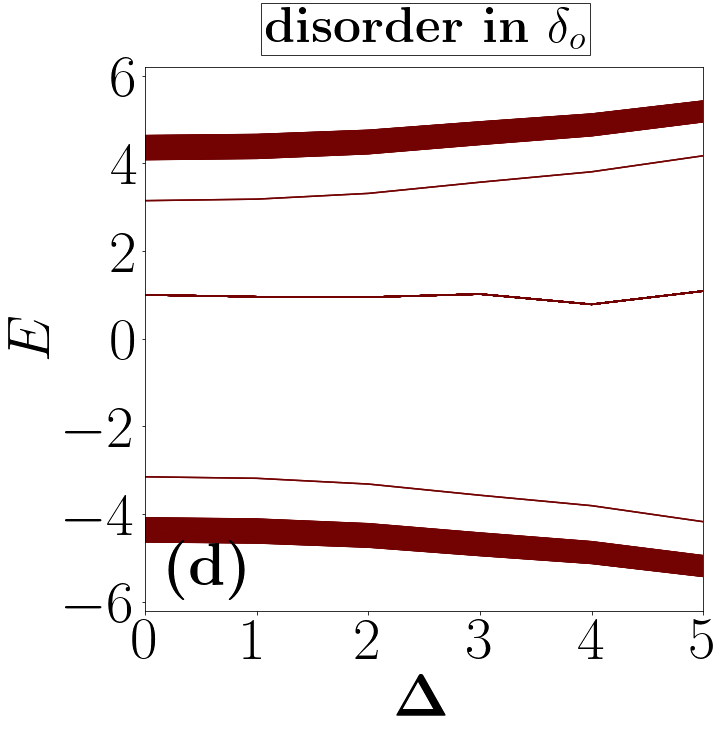}   
        \label{fig:sub4}
    \end{subfigure}    
    \caption{\CH{The energy spectrum versus the perturbation disorder strength of disordered Eqs.~\Crefrange{eq:eq9}{eq:eq12} at a fixed $J_{1} = 0.2$ and $J_{2} = 3$ for $N = 50$ unit cells under OBC. The starting point for each case is $\alpha_{o}, \beta_{o}, \gamma_{o}, \delta_{o} = 1$ respectively. (a) Disorder acts on $\alpha_{o}$, (b) Disorder acts on  $\beta_{o}$, (c) Disorder acts on $\gamma_{o}$, and (d) Disorder acts on $\delta_{o}$. Each data point is averaged over 200 iterations.}}
    \label{fig:fig11dis}
\end{figure}

\begin{figure}[h]
    \centering
    \begin{subfigure}[b]{0.23\textwidth}
        \centering
        \includegraphics[width=\textwidth]{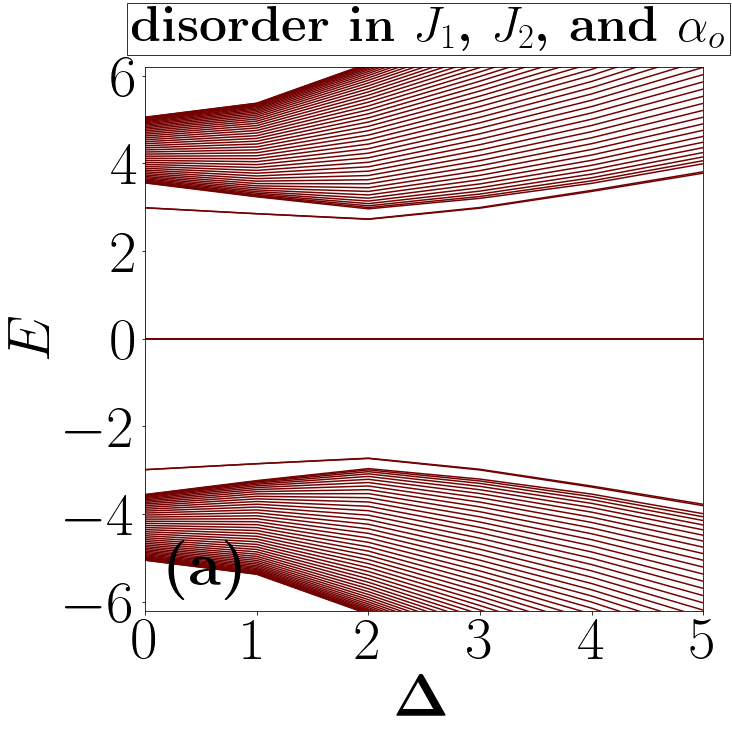}
        \label{fig:sub1}
    \end{subfigure}
    \begin{subfigure}[b]{0.23\textwidth}
        \centering
        \includegraphics[width=\textwidth]{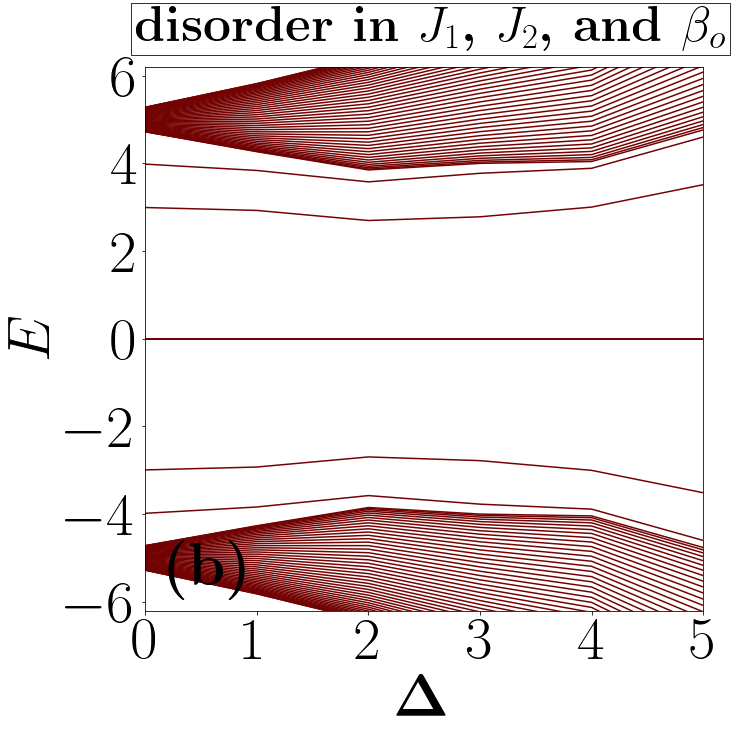}      
        \label{fig:sub2}
    \end{subfigure}   
    \begin{subfigure}[b]{0.23\textwidth}
        \centering
        \includegraphics[width=\textwidth]{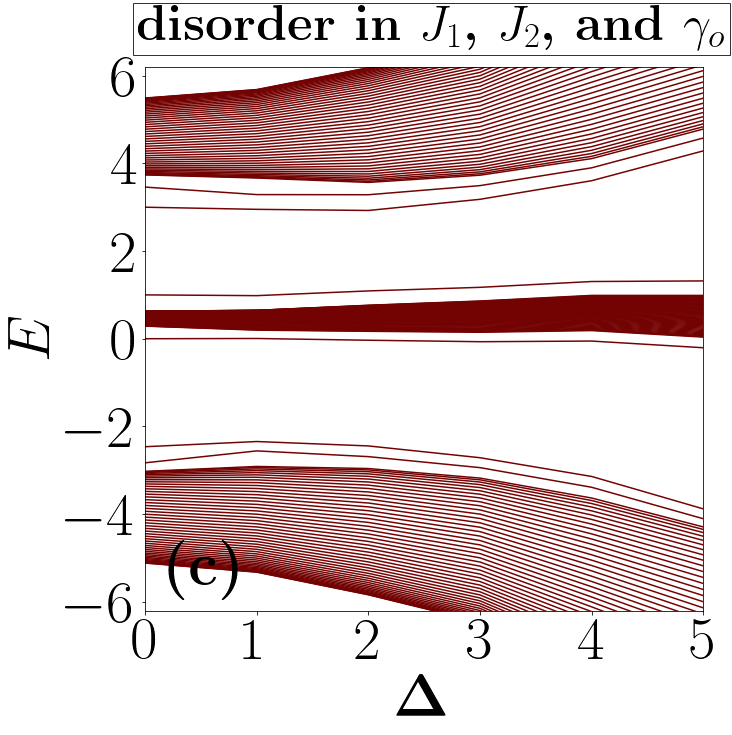}
        \label{fig:sub3}
    \end{subfigure}
    \begin{subfigure}[b]{0.23\textwidth}
        \centering
        \includegraphics[width=\textwidth]{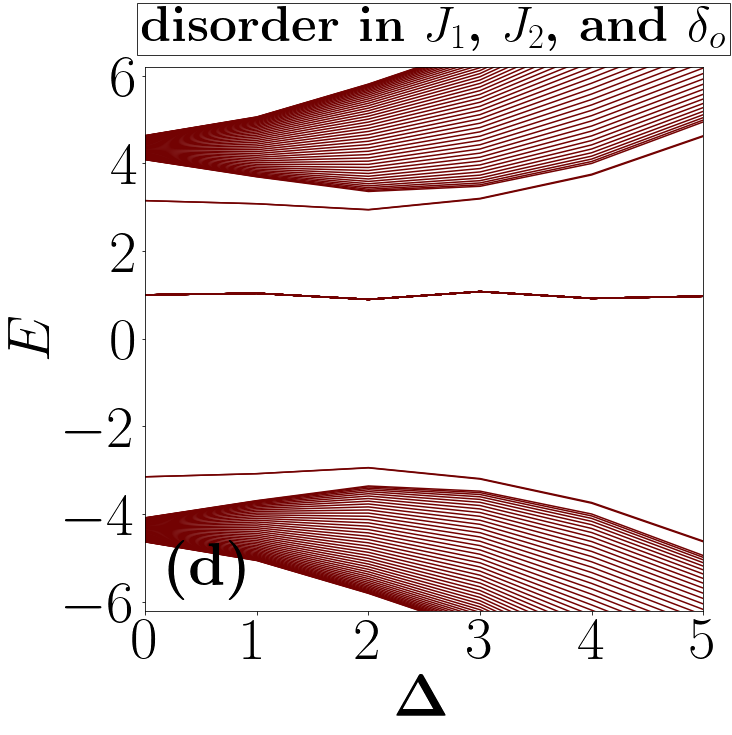}   
        \label{fig:sub4}
    \end{subfigure}    
    \caption{\CH{The same as in Fig.~\ref{fig:fig11dis} but disorder also acts on both hopping parameters.}}
    \label{fig:fig20}
\end{figure}


\CH{In Figs.~\ref{fig:fig11dis} and ~\ref{fig:fig20}, we further examine the fate of the system's edge states in the simultaneous presence of perturbations considered in Sec.~\ref{pert} ($\alpha_{o}, \beta_{o}, \gamma_{o},$ and $\delta_{o}$) and spatial disorder. To yield a comprehensive analysis, two cases have been considered in our calculations. In Fig.~\ref{fig:fig11dis}, we only include the presence of disorder in the perturbation parameters, so as to understand the individual effect of disorder on the previously studied perturbations. In Fig.~\ref{fig:fig20}, disorder is incorporated in all system parameters (perturbations and both hopping parameters), so as to obtain the overall effect of disorder in a more physical setting. In all cases, the edge states' behaviours unveiled in Sec.~\ref{pert} remain intact even at large disorder, thereby highlighting the robustness of the obtained edge states.}


\section{Concluding remarks}
\label{conc}

We have presented a trimerized extension to the paradigmatic SSH model and uncovered its rich spectral and topological features. Unlike similar extensions proposed in existing literature, our model was obtained by replacing the sublattice Pauli matrices in the regular SSH model by their three-dimensional counterparts. Consequently, our trimer SSH models were shown to demonstrate fundamentally different edge state behaviors and symmetry protection. For example, our model possesses time-reversal, chiral, particle-hole, and inversion symmetries. Moreover, in the topologically nontrivial regime, two pairs of edge states emerge symmetrically at positive and negative energies. In the presence of perturbations, such edge states remain robust provided the time-reversal, chiral, and particle-hole symmetries are preserved. \CH{We have further demonstrated our model's considerable robustness against spatial disorder, as expected from a topological phase.}

While the edge states remain robust in the presence of symmetry-preserving perturbations, their structure may change. Remarkably, this could in turn lead to an interesting scenario in which edge states are localized only on one edge of the system, as we explicitly demonstrated with our choice of perturbation $\beta(k)$ in Sec.~\ref{pert} (see also Fig.~\ref{fig:fig11} in Appendix~\ref{app:B}). Such a feature could find some useful applications in the area of quantum communications, particularly related to the task of quantum state transfers \cite{Zurita2023, Chang2023, Zheng2023, C.Wang2022, Palaiodimopoulos2021, D'Angelis2020, Tan2020, Mei2018}. Indeed, transferring some quantum information from one edge of the lattice to the other could be accomplished by first encoding it in the subspace spanned by the vacuum state and the edge states that are originally localized solely on one edge, then slowly transforming the system's Hamiltonian into that which supports edge states solely on the other edge. As long as quantum adiabaticity holds, it is then guaranteed that the quantum information is now encoded in the subspace spanned by the vacuum state and the edge states localized solely on the other edge, thereby completing the intended transfer. \CH{It should however be noted that some care must be taken in investigating such a quantum state transfer via quantum dots implementation of our model, as there might be additional degrees of freedom that must be taken into account \cite{O2017}. In this case, The quantum state transfer application of our model is beyond the scope of this work and is best left for potential follow-up work.}

In the future, it would be worthwhile to extend our procedure for obtaining a family of extended SSH models with $n$ sites per unit cell. Such models are expected to exhibit even richer properties, the full analysis of which deserves a separate study. \CH{However, even within the same model studied in this paper, interesting novel phenomena could still arise when subjecting the system to more sophisticated boundary conditions, e.g., those that form M\"{o}bius strip in the spirit of Ref.~\cite{Huang11}.} Another interesting direction to pursue is to investigate the effect of periodic driving on the present model. In particular, periodically driving a regular SSH model has already been shown to yield novel topological features with no static counterparts \cite{Cheng2019, Z.Cheng2023}. It is thus envisioned that periodically driving our trimer SSH model will lead to more unexpected topological phenomena. Finally, non-Hermiticity \cite{Jangjan2024, Yao2018, Lee2016, Lieu2018, Wu2021, Halder2023, Okuma2023}, nonlinearity \cite{Jezequel2022, Y.Ma2021, Tuloup2020}, and interaction effect \cite{Koor2022, Feng2022, Yu2020, Melo2023} are other aspects that could be considered to further enrich the physics of our model.

 \begin{acknowledgements}
 This work was supported by the Deanship of Research
Oversight and Coordination (DROC) at King Fahd University of Petroleum \& Minerals (KFUPM) through project No.~EC221010.
 \end{acknowledgements}

\appendix
\section{Detailed analytical calculation of the normalized sublattice Zak's phase}
\label{app:0}

Recall that the normalized sublattice Zak's phase is given as
\begin{equation}
Z^{\zeta}_{\lambda} = \frac{i}{2}\oint dk \langle \widetilde{\psi}^{\zeta}_{\lambda}(k) | \partial_{k} \widetilde{\psi}^{\zeta}_{\lambda}(k) \rangle,
\end{equation}
where
\[ |\widetilde{\psi}^{\zeta}_{\lambda}(k) \rangle = \frac{P_{\zeta} | \psi_{\lambda}(k) \rangle}{\sqrt{\langle \psi_{\lambda}(k)| P_{\zeta} | \psi_{\lambda}(k) \rangle}} , \]
and the eigenstates for $\Gamma$ are
\[ |A \rangle = \begin{pmatrix} 1 \\ 0 \\ 0 \end{pmatrix}, \quad |B \rangle = \begin{pmatrix} 0 \\ 1 \\ 0 \end{pmatrix}, \quad |C \rangle = \begin{pmatrix} 0 \\ 0 \\ 1 \end{pmatrix} . \]
We will now compute $Z^{\zeta}_{\lambda}$ for all possible values of $\zeta$ and $\lambda$.\\

{\it The case $\zeta=A$}:\\
\noindent The eigenstate corresponding to $E_{1} = 0$ is

\begin{align*}
|\psi_{1}(k) \rangle &= \begin{pmatrix}
\frac{-(J_1 + J_2\,e^{-i k})}{J_1 + J_2\,e^{i k}} \\
\\
0 \\
\\
0
\end{pmatrix}.
\end{align*}

\[ P_{A} = \begin{pmatrix} 1 & 0 & 0 \\ 0 & 0 & 0 \\ 0 & 0 & 0  \end{pmatrix}, \]

\[ P_{A}| \psi_{1}(k) \rangle = \begin{pmatrix}
\frac{-(J_1 + J_2\,e^{-i k})}{J_1 + J_2\,e^{i k}} \\
\\
0 \\
\\
0
\end{pmatrix}, \]

so that

\[ \langle \psi_{1}(k) | P_{A} |\psi_{1}(k) \rangle = 1,\]

\[ | \widetilde{\psi}^{A}_{1}(k) \rangle = \begin{pmatrix}
\frac{-(J_1 + J_2\,e^{-i k})}{J_1 + J_2\,e^{i k}} \\
\\
0 \\
\\
0
\end{pmatrix}, \]

\begin{align*}
\partial_{k}| \widetilde{\psi}^{A}_{1}(k) \rangle &= \begin{pmatrix}
2iJ_{2}\left\{ \frac{J_2 + J_1\,\cos(k)}{(J_1+J_2\,e^{ik})^2}  \right\} \\
\\
0 \\
\\
0
\end{pmatrix} .
\end{align*}
Finally, straightforward calculation yields

\begin{align*}
Z^{A}_{1} &= \frac{i}{2}\oint dk \langle \widetilde{\psi}^{A}_{1}(k) |\partial_{k} \widetilde{\psi}^{A}_{1}(k) \rangle \\
&= \frac{i}{2}\oint dk \left\{ (-2\,i\,J_2)\,\frac{J_2 + J_1\,\cos(k)}{J_1^2 + J_2^2 + 2\,J_1\,J_2\,\cos(k)} \right\}\\
&= \oint dk \left\{ J_2\,\frac{J_2 + J_1\,\cos(k)}{J_1^2 + J_2^2 + 2\,J_1\,J_2\,\cos(k)} \right\}\\ &= 
\begin{cases}
    2\pi, & \text{if } J_2 > J_1 \\
    0\phantom{\pi},   & \text{if } J_2 < J_1
\end{cases}.
\end{align*}

The eigenstates corresponding to \[ E_{2, 3} = \pm \sqrt{2(J_{1}^{2} + J_{2}^{2} + 2 J_{1} J_{2} \cos(k))} \] are
\begin{align*}
|\psi_{2,3}(k) \rangle &= \begin{pmatrix}
\frac{J_1 + J_2\,e^{-i k}}{J_1 + J_2\,e^{i k}} \\
\\
\pm \frac{\sqrt{2(J_1^2 + J_2^2 + 2 J_1\,J_2\,\cos(k))}}{J_1 + J_2\,e^{i k}} \\
\\
0
\end{pmatrix}.
\end{align*}

\[ P_{A} |\psi_{2,3}(k) \rangle_ = \begin{pmatrix}
\frac{J_1 + J_2\,e^{-i k}}{J_1 + J_2\,e^{i k}}\\
\\
0 \\
\\
0
\end{pmatrix},
\]
and
\[ \langle \psi_{2,3}(k) | P_{A} |\psi_{2,3}(k) \rangle = 1,\]
so that
\[ | \widetilde{\psi}^{A}_{2,3}(k) \rangle = \begin{pmatrix}
\frac{J_1 + J_2\,e^{-i k}}{J_1 + J_2\,e^{i k}} \\
\\
0 \\
\\
0
\end{pmatrix}, \]

\begin{align*}
\partial_{k}| \widetilde{\psi}^{A}_{2,3}(k) \rangle &= \begin{pmatrix}
-2iJ_{2}\left\{ \frac{J_2 + J_1\,\cos(k)}{(J_1+J_2\,e^{ik})^2}  \right\} \\
\\
0 \\
\\
0
\end{pmatrix} .
\end{align*}

Finally, straightforward calculation yields
\begin{align*}
Z^{A}_{2,3} &= \frac{i}{2}\oint dk \langle \widetilde{\psi}^{A}_{2,3}(k) |\partial_{k} \widetilde{\psi}^{A}_{2,3}(k) \rangle \\
&= \frac{i}{2}\oint dk \left\{ (-2\,i\,J_2)\,\frac{J_2 + J_1\,\cos(k)}{J_1^2 + J_2^2 + 2\,J_1\,J_2\,\cos(k)} \right\}\\
&= \oint dk \left\{ J_2\,\frac{J_2 + J_1\,\cos(k)}{J_1^2 + J_2^2 + 2\,J_1\,J_2\,\cos(k)} \right\}\\ &= \begin{cases}
    2\pi, & \text{if } J_2 > J_1 \\
    0\phantom{\pi},   & \text{if } J_2 < J_1
\end{cases}.
\end{align*}
As the name suggests, the normalized sublattice Zak's phase is defined modulo $2\pi$. Therefore, the above results imply that $Z_\ell^{A}=0$ for all $\ell=1,2,3$.\\ 

{\it The case $\zeta=B$}:\\
\noindent The eigenstate corresponding to $E_{1} = 0$ is
\begin{align*}
|\psi_{1}(k) \rangle &= \begin{pmatrix}
\frac{-(J_1 + J_2\,e^{-i k})}{J_1 + J_2\,e^{i k}} \\
\\
0 \\
\\
0
\end{pmatrix}.
\end{align*}

\[ P_{B} = \begin{pmatrix} 0 & 0 & 0 \\ 0 & 1 & 0 \\ 0 & 0 & 0  \end{pmatrix}, \]

\[ P_{B}| \psi_{1}(k) \rangle = \begin{pmatrix}
0 \\
\\
0 \\
\\
0
\end{pmatrix}, \]

this leads to
\begin{align*}
\langle \widetilde{\psi}^{B}_{1}(k) |\partial_{k} \widetilde{\psi}^{B}_{1}(k) \rangle = 0,
\end{align*}
and therefore,
\begin{align*}
Z^{B}_{1} &= \frac{i}{2}\oint dk \langle \widetilde{\psi}^{B}_{1}(k) |\partial_{k} \widetilde{\psi}^{B}_{1}(k) \rangle \\
&= 0.
\end{align*}

The eigenstates corresponding to \[ E_{2, 3} = \pm \sqrt{2(J_{1}^{2} + J_{2}^{2} + 2 J_{1} J_{2} \cos(k))} \] are
\begin{align*}
|\psi_{2,3}(k) \rangle &= \begin{pmatrix}
\frac{J_1 + J_2\,e^{-i k}}{J_1 + J_2\,e^{i k}} \\
\\
\pm \frac{\sqrt{2(J_1^2 + J_2^2 + 2 J_1 J_2 \cos(k))}}{J_1 + J_2\,e^{i k}} \\
\\
0
\end{pmatrix}.
\end{align*}
It follows that

\[ P_{B} |\psi_{2,3}(k) \rangle_ = \begin{pmatrix}
0 \\
\\
\pm \frac{\sqrt{2(J_1^2 + J_2^2 + 2 J_1 J_2 \cos(k))}}{J_1 + J_2\,e^{i k}} \\
\\
0
\end{pmatrix},
\]
and
\[ \langle \psi_{2,3}(k) | P_{B} |\psi_{2,3}(k) \rangle = 2,\]
so that
\[ | \widetilde{\psi}^{B}_{2, 3}(k) \rangle = \begin{pmatrix}
0 \\
\\
\pm \frac{\sqrt{(J_1^2 + J_2^2 + 2 J_1 J_2 \cos(k))}}{J_1 + J_2\,e^{i k}} \\
\\
0
\end{pmatrix}, \]
and
\begin{align*}
\partial_{k}| \widetilde{\psi}^{B}_{2, 3}(k) \rangle &= \begin{pmatrix}
0 \\
\\
\mp \left\{ \frac{J_1\,J_2\,\sin(k)}{(J_1+J_2\,e^{ik})\sqrt{J_1^2 + J_2^2 + 2\,J_1\,J_2\,\cos(k)}} \right. & \\
\quad + \left. \frac{i\,J_2\, e^{ik}\sqrt{J_1^2 + J_2^2 + 2 J_1 J_2 \cos(k)}}{(J_1 + J_2\,e^{ik})^2} \right\} \\
\\
0
\end{pmatrix} .
\end{align*}
Finally, straightforward calculation yields
\begin{align*}
Z^{B}_{2, 3} &= \frac{i}{2}\oint dk \langle \widetilde{\psi}^{B}_{2,3}(k) |\partial_{k} \widetilde{\psi}^{B}_{2,3}(k) \rangle \\
&= - \frac{i}{2}\oint dk \left\{ \frac{J_1\,J_2\,\sin(k)}{J_1^2 + J_2^2 + 2\,J_1\,J_2\,\cos(k)} + i\frac{J_2 e^{ik}}{J_1 +J_2\,e^{ik}} \right\}\\
&= \begin{cases}
    \pi, & \text{if } J_2 > J_1 \\
    0,   & \text{if } J_2 < J_1
\end{cases}.
\end{align*}
As reported in the main text. \\

{\it The case $\zeta=C$}:\\
\[ P_{C} = \begin{pmatrix} 0 & 0 & 0 \\ 0 & 0 & 0 \\ 0 & 0 & 1  \end{pmatrix}, \]

\[ P_{C}| \psi_{1, 2, 3}(k) \rangle = \begin{pmatrix}
0 \\
\\
0 \\
\\
0
\end{pmatrix}, \]

\[ \langle \psi_{1,2,3}(k) | P_{C} |\psi_{1,2,3}(k) \rangle = 0,\]

\[ | \widetilde{\psi}^{C}_{1, 2, 3}(k) \rangle = \begin{pmatrix}
0 \\
\\
0 \\
\\
0
\end{pmatrix}, \]

\begin{align*}
\partial_{k}| \widetilde{\psi}^{C}_{1, 2, 3}(k) \rangle &= \begin{pmatrix}
0  \\
\\
0 \\
\\
0
\end{pmatrix},
\end{align*}
this leads to
\begin{align*}
\langle \widetilde{\psi}^{C}_{\lambda}(k) |\partial_{k} \widetilde{\psi}^{C}_{\lambda}(k) \rangle = 0,
\end{align*}
and finally
\begin{align*}
Z^{C}_{\lambda} &= \frac{i}{2}\oint dk \langle \widetilde{\psi}^{C}_{\lambda}(k) |\partial_{k} \widetilde{\psi}^{C}_{\lambda}(k) \rangle \\
&= 0.
\end{align*}

\section{Detailed analytical calculation of the edge states at $J_1=0$} 
\label{app:A}

By inspecting Fig.~\ref{fig:fig3} in the main text, it is clear that at $J_1=0$, the edge states correspond to energy $E=\pm J_2$. To analytically obtain the exact form of the edge states at this special parameter value, we thus attempt to solve
\begin{align}
\mathcal{H} \ket{\psi} &= \pm J_{2} \ket{\psi},
\label{eq:eqB1}
\end{align}
where Eq.~(\ref{eq:eq6}) becomes (under $J_1=0$), 
\begin{flalign}
 \mathcal{H} &= \sum_{j=1}^{N-1}\, \left( J_{2}\,| A,j+1 \rangle \langle B,j | + J_{2}\,| B,j+1 \rangle \langle C,j | \right) + \text{\textit{h.c.}} ,
 \label{eq:eqB2}
\end{flalign}
Using the same notation as that in Sec.~\ref{edge}, we first write 
\begin{flalign}
\ket{\psi} &= \sum_{j=1}^{N}\,  a_{j}\,| A,j \rangle  + b_{j}\,| B,j \rangle  + c_{j}\,| C,j \rangle,
\label{eq:eqB3}
\end{flalign}
By substituting Eq.~(\ref{eq:eqB2}) and Eq.~(\ref{eq:eqB3}) into Eq.~(\ref{eq:eqB1}) with $E=J_2$,

\begin{align*}
a_1 &= 0 ,\\
J_2\,a_2 &= J_2 \,b_1 , \\
\\
J_2\,b_2 &= J_2 \,c_1 ,\\
J_2\,b_1 &= J_2 \,a_2 ,\\
J_2\,c_1 + J_2\,a_3 &= J_2\,b_2 , \\
\\
J_2\,b_3 &= J_2 \,c_2 ,\\
J_2\,b_2 &= J_2 \,a_3 ,\\
J_2\,c_2 + J_2\,a_4 &= J_2\,b_3 , 
\\
& \vdots 
\\
J_2\,b_j &= J_2 \,c_{j-1} ,\\
J_2\,b_{j-1} &= J_2 \,a_{j+1} ,\\
J_2\,c_{j-1} + J_2\,a_{j+1} &= J_2\,b_{j} ,\\
\end{align*}

By solving these equations:
\begin{align*}
b_{j} &= a_{j+1} \\
c_j &= b_{j+1} = a_{j+2} \\
c_{j} + a_{j+2} &= b_{j+1}
\end{align*}
The last two equations imply $c_j = b_{j+1}=a_{j+2}=0$. Consequently, the only nonzero elements are $b_1=a_2$. For $E=-J_2$, a similar calculation yields $b_1=-a_2$ and all other elements being zero. Plugging these results to Eq.~(\ref{eq:eqB3}), the left localized edge states are obtained as
\[
\ket{\psi_{L,\pm}} =
|B,1 \rangle \pm |A,2 \rangle.
\]
By repeating the calculation above from $c_N$ going backwards, we obtain similar expressions for the right localized edge states, i.e.,
\[
\ket{\psi_{R,\pm}} =
|B,N \rangle \pm |C,N-1 \rangle. 
\]

\section{Wave function profiles of the edge states in the presence of perturbations} 
\label{app:B}

\begin{figure*}[htpb]
    \centering
    \begin{minipage}[b]{0.75\textwidth}
        \centering
        \includegraphics[width=\textwidth]{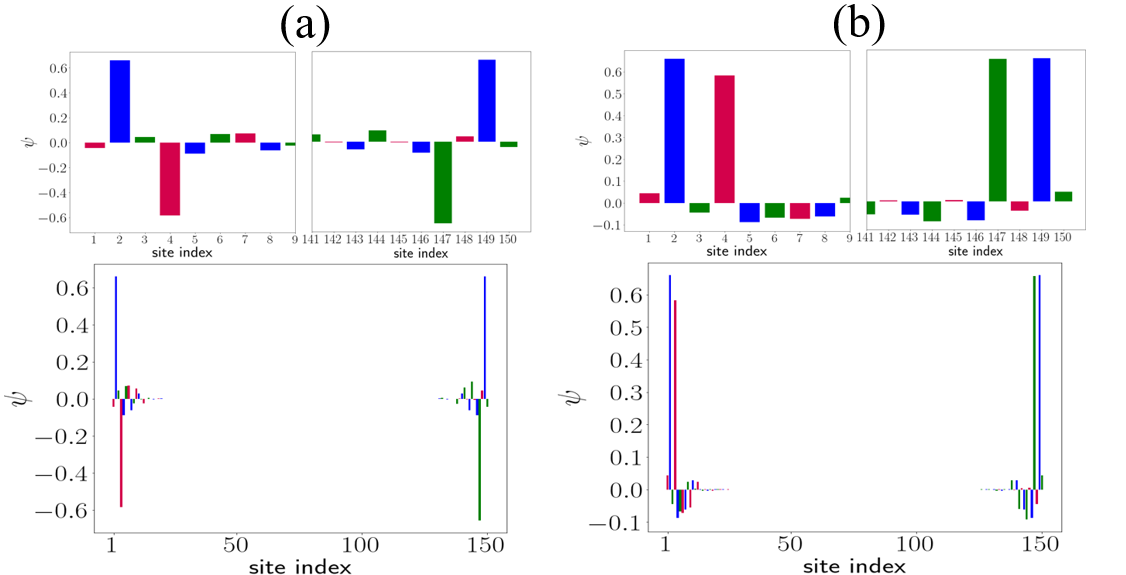}
    \end{minipage}
    \caption{Wave function profiles of the system's edge states in the presence of $\alpha$ perturbation (see main text) at $\alpha_{\text{o}}=1$ and $N=50$ unit cells. Subplot (a) for $E \approx - 2.99$ and subplot (b) for $E \approx 2.99$. The upper panels show the zoomed-in view of each edge state near the appropriate (left or right) lattice edge.}
    \label{fig:fig7}
\end{figure*}

\begin{figure*}[htpb]
    \centering
    \begin{subfigure}[b]{0.75\textwidth}
        \centering
        \includegraphics[width=\textwidth]{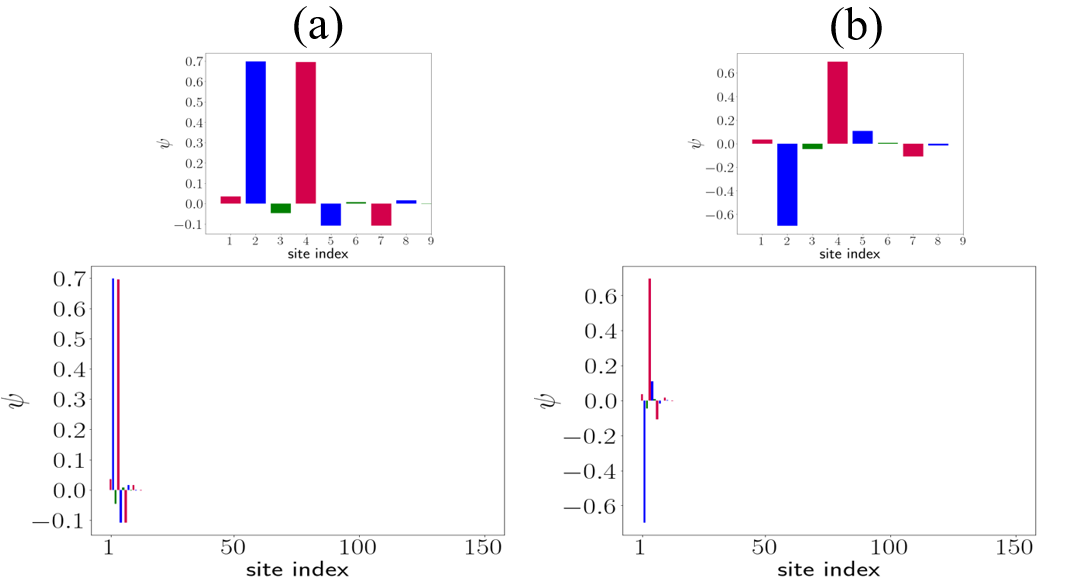}
        \caption*{\textbf{}}
        \label{fig:sub1}
    \end{subfigure}
    \hspace{0.00\textwidth}
    \begin{subfigure}[b]{0.75\textwidth}
        \centering
        \includegraphics[width=\textwidth]{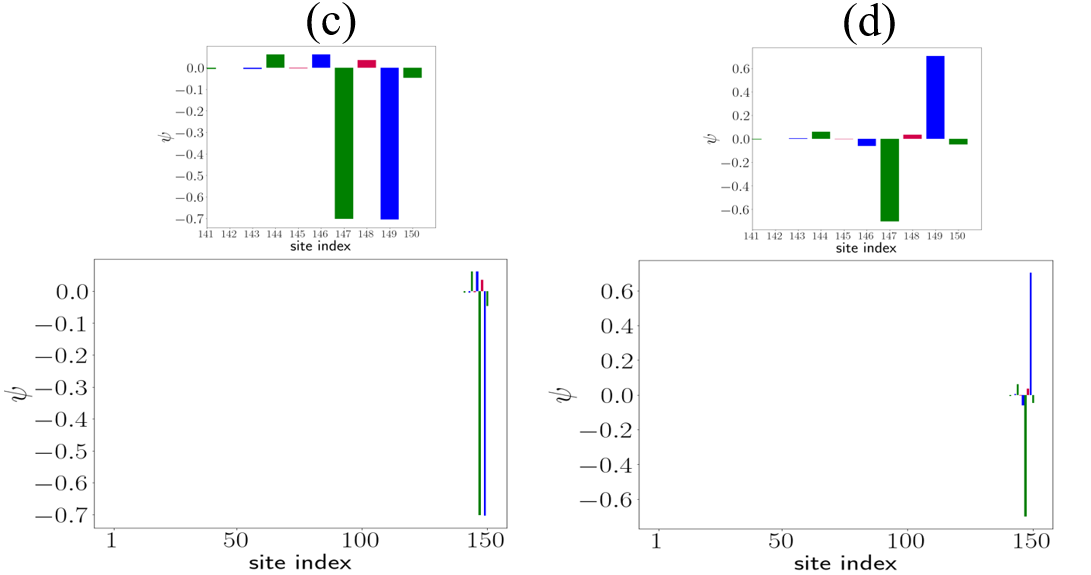}
        \caption*{\textbf{}}
        
        \label{fig:sub4}
    \end{subfigure}
    \caption{Wave function profiles of the system's edge states in the presence of $\beta$ perturbation (see main text) at $\beta_{\text{o}}=1$ and $N=50$ unit cells. Subplot (a) at $E \approx 3.98$, subplot (b) at $E \approx -3.98$, subplot (c) at $E \approx 2.99$, and subplot (d) at $E \approx -2.99$. The upper panels show the zoomed-in view of each edge state near the appropriate (left or right) lattice edge.}
    \label{fig:fig8}
\end{figure*}

\begin{figure*}[htpb]
    \centering
    \begin{subfigure}[b]{0.75\textwidth}
        \centering
        \includegraphics[width=\textwidth]{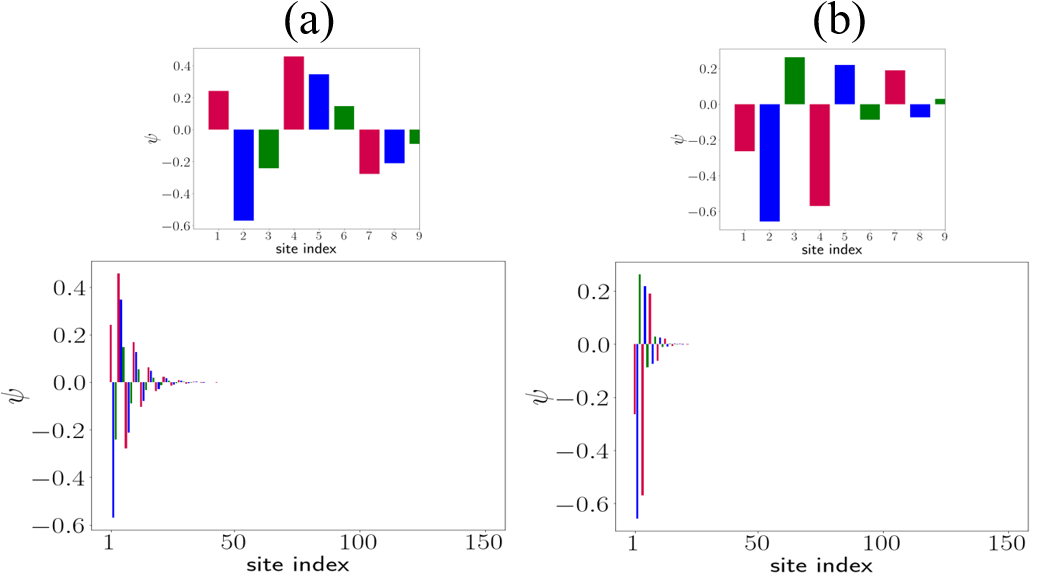}
        \caption*{\textbf{}}
        \label{fig:sub1}
    \end{subfigure}
    \hspace{0.0\textwidth}
    \begin{subfigure}[b]{0.75\textwidth}
        \centering
        \includegraphics[width=\textwidth]{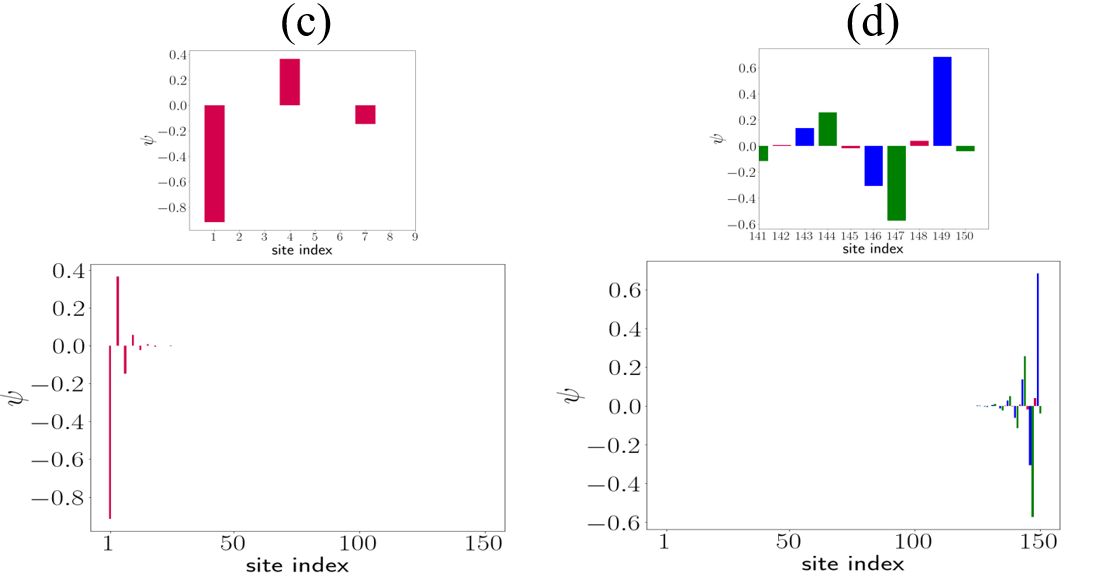}
        \caption*{\textbf{}}
        \label{fig:sub2}
    \end{subfigure}
    \hspace{0.0\textwidth}
    \begin{subfigure}[b]{0.75\textwidth}
        \centering
        \includegraphics[width=\textwidth]{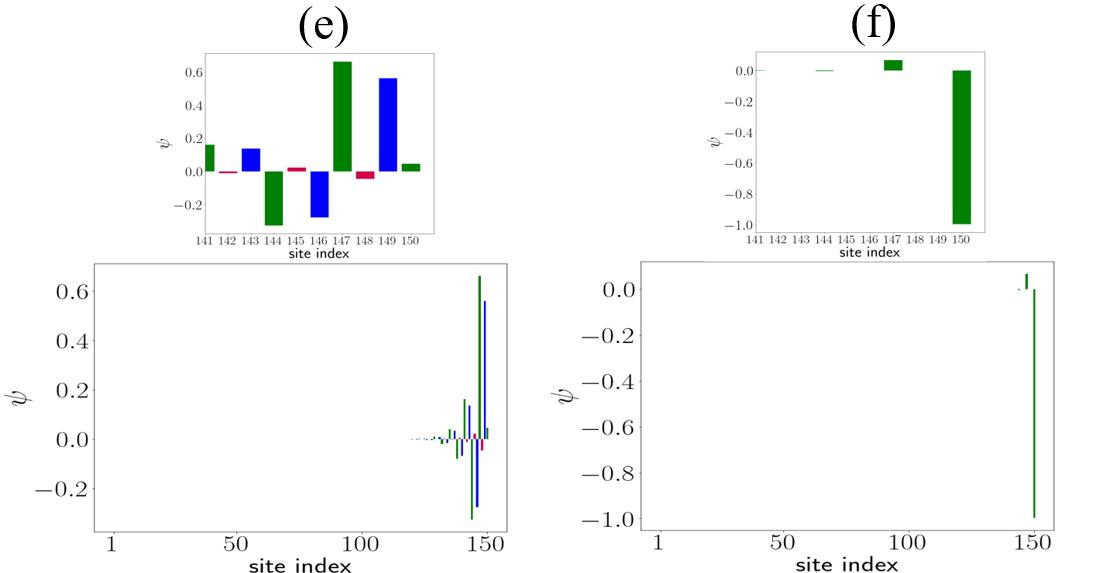}
        \caption*{\textbf{}}
        \label{fig:sub3}
    \end{subfigure}
        \label{fig:sub6}
    \caption{Wave function profiles of the system's edge states in the presence of $\gamma$ perturbation (see main text) at $\gamma_{\text{o}}=1$ and $N=50$ unit cells. Subplot (a) at $E \approx -2.83$, subplot (b) at $E \approx 3$, subplot (c) at $E \approx 0$, subplot (d) at $E \approx -2.46$, subplot (e) at $E \approx 3.46$, and subplot (f) at $E \approx 1$. The upper panels show the zoomed-in view of each edge state near the appropriate (left or right) lattice edge.}
    \label{fig:fig9}
\end{figure*}

\begin{figure*}
    \centering
    
    \begin{subfigure}[b]{0.75\textwidth}
        \centering
        \includegraphics[width=\textwidth]{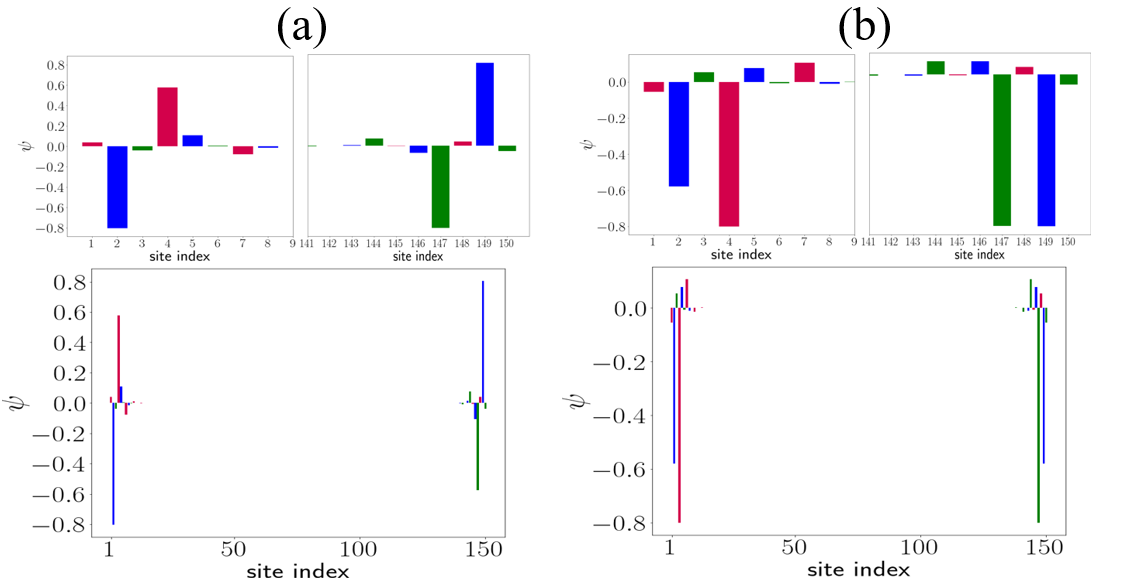}
        \caption*{\textbf{}}
        \label{fig:sub1}
    \end{subfigure}
    \hspace{0.0\textwidth}
    \begin{subfigure}[b]{0.37\textwidth}
        \centering
        \includegraphics[width=\textwidth]{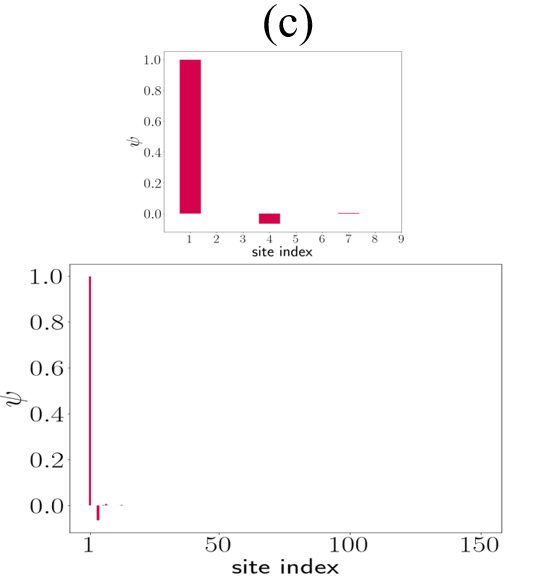}
        \caption*{\textbf{}}
        \label{fig:sub2}
    \end{subfigure}
    
    \caption{Wave function profiles of the system's edge states in the presence of $\delta$ perturbation (see main text) at $\delta_{\text{o}}=1$ and $N=50$ unit cells. Subplot (a) at $E \approx 3.15$, subplot (b) at $E \approx -3.15$, and subplot (c) at $E \approx 1$. The upper panels show the zoomed-in view of each edge state near the appropriate (left or right) lattice edge.}
    \label{fig:fig10}
\end{figure*}

\begin{figure*}
    \centering
    
    \begin{subfigure}[b]{0.75\textwidth}
        \centering
        \includegraphics[width=\textwidth]{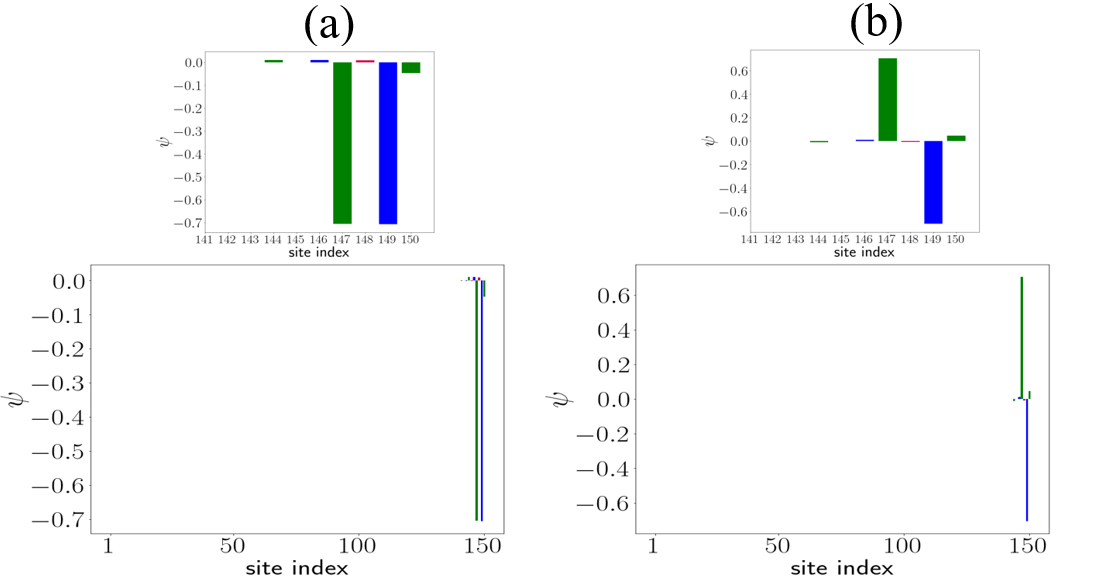}
        \caption*{\textbf{}}
        \label{fig:sub1}
    \end{subfigure}
    \hspace{0.0\textwidth}
    \begin{subfigure}[b]{0.37\textwidth}
        \centering
        \includegraphics[width=\textwidth]{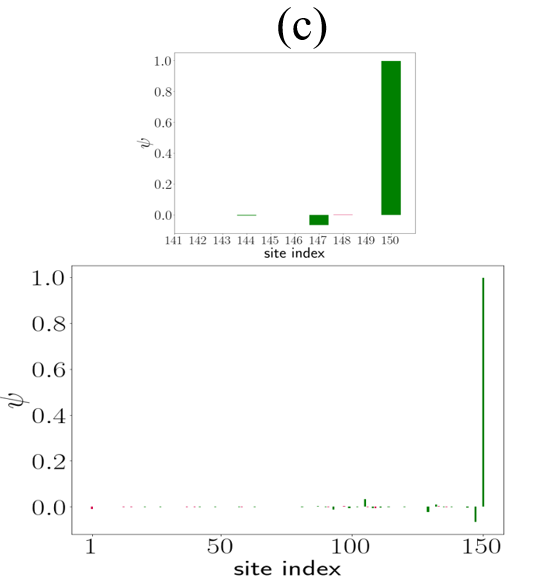}
        \caption*{\textbf{}}
        \label{fig:sub2}
    \end{subfigure}
 
    \caption{Wave function profiles of the system's edge states in the presence of $\beta$ perturbation at large $\beta_{\text{o}}=13$ value and $N=50$ unit cells. Subplot (a) at $E \approx 3$, subplot (b) at $E \approx -3$, and subplot (c) at $E \approx 0$. The upper panels show the zoomed-in view of each edge state near the right lattice edge.}
    \label{fig:fig11}
\end{figure*}

In the main text, we have demonstrated the robustness of the system's edge states in the presence of symmetry-preserving perturbations. For completeness, we present in this section the wave function profiles corresponding to the surviving edge states under the perturbations considered in the main text. Our results are summarized in Figs.~\ref{fig:fig7}-\ref{fig:fig10}. Observe that in all cases, the edge states that originate from the unperturbed scenario (whose main peaks are at $|B,1\rangle \pm |A,2\rangle$ or $|B, N\rangle \pm |C, N-1\rangle$) remain present, thus demonstrating their robustness. Finally, Fig.~\ref{fig:fig11} demonstrates an instance in which the system's edge states are solely localized on the right edge when the perturbation $\beta_{\text{o}}$ is large.

\begin{thebibliography}{0}%
\makeatletter
\providecommand \@ifxundefined [1]{%
 \@ifx{#1\undefined}
}%
\providecommand \@ifnum [1]{%
 \ifnum #1\expandafter \@firstoftwo
 \else \expandafter \@secondoftwo
 \fi
}%
\providecommand \@ifx [1]{%
 \ifx #1\expandafter \@firstoftwo
 \else \expandafter \@secondoftwo
 \fi
}%
\providecommand \natexlab [1]{#1}%
\providecommand \enquote  [1]{``#1''}%
\providecommand \bibnamefont  [1]{#1}%
\providecommand \bibfnamefont [1]{#1}%
\providecommand \citenamefont [1]{#1}%
\providecommand \href@noop [0]{\@secondoftwo}%
\providecommand \href [0]{\begingroup \@sanitize@url \@href}%
\providecommand \@href[1]{\@@startlink{#1}\@@href}%
\providecommand \@@href[1]{\endgroup#1\@@endlink}%
\providecommand \@sanitize@url [0]{\catcode `\\12\catcode `\$12\catcode `\&12\catcode `\#12\catcode `\^12\catcode `\_12\catcode `\%12\relax}%
\providecommand \@@startlink[1]{}%
\providecommand \@@endlink[0]{}%
\providecommand \url  [0]{\begingroup\@sanitize@url \@url }%
\providecommand \@url [1]{\endgroup\@href {#1}{\urlprefix }}%
\providecommand \urlprefix  [0]{URL }%
\providecommand \Eprint [0]{\href }%
\providecommand \doibase [0]{http://dx.doi.org/}%
\providecommand \selectlanguage [0]{\@gobble}%
\providecommand \bibinfo  [0]{\@secondoftwo}%
\providecommand \bibfield  [0]{\@secondoftwo}%
\providecommand \translation [1]{[#1]}%
\providecommand \BibitemOpen [0]{}%
\providecommand \bibitemStop [0]{}%
\providecommand \bibitemNoStop [0]{.\EOS\space}%
\providecommand \EOS [0]{\spacefactor3000\relax}%
\providecommand \BibitemShut  [1]{\csname bibitem#1\endcsname}%
\let\auto@bib@innerbib\@empty
\end{thebibliography}%


\clearpage
\begin{thebibliography}{99}

        
        \bibitem{Thouless82} D.~J.~Thouless, M.~Kohmoto, M.~P.~Nightingale, and M.~den Nijs, Phys.~Rev.~Lett.~{\bf 49}, 405 (1982).

        \bibitem{Wen90} X.-G.~Wen, Int.~J.~of Mod.~Phys.~B~{\bf 4}, 239-271 (1990).

        \bibitem{Hasan2010} M.~Z.~Hasan and C.~L.~Kane, Rev.~Mod.~Phys.~{\bf 82}, 3045 (2010).

        \bibitem{He2022} Q.~L.~He, T.~L.~Hughes, N.~P.~Armitage, Y.~Tokura, and K.~L.~Wang, N.~Materials {\bf 21}, 15–23 (2022).

        \bibitem{He2019} M.~He, H.~Sun, and Q.~L.~He, Front.~Phys.~{\bf 14}, 43401 (2019).

        \bibitem{Fan2016} Y.~Fan and K.~L.~Wang, SPIN {\bf 6}, 1640001 (2016).
        
        \bibitem{Yang2023} Y.~Yang, H.~Sun, J.~Lu, X.~Huang, W.~Deng, and Z.~Liu, Communications Physics {\bf 6}, 143 (2023).

        \bibitem{Cai2023} L.~Cai, R.~Li, X.~Wu, B.~Huang, Y.~Dai, and C.~Niu, Phys.~Rev.~B~{\bf 107}, 245116 (2023).

        \bibitem{Zhou2022} L.~Zhou, R.~W.~Bomantara, and S.~Wu, SciPost~Phys.~{\bf 13}, 015 (2022).

        \bibitem{Denner2021} M.~M.~Denner, A.~Skurativska, F.~Schindler, M.~H.~Fischer, R.~Thomale, T.~Bzdušek, and T.~Neupert, Nat.~Commun.~{\bf 12}, 5681 (2021).
        
        \bibitem{Tokura2019} Y.~Tokura, K.~Yasuda, and A.~Tsukazaki, Nat.~Rev.~Phys.~{\bf 1}, 126–143 (2019).

        \bibitem{Su1980} W.~P.~Su, J.~R.~Schrieffer, and A.~J.~Heeger, Phys.~Rev.~B~{\bf 22}, 2099 (1980).

        \bibitem{Asboth2016} J.~K.~Asb\'{o}th, L.~Oroszlány, and A.~Pályi, Lecture Notes in Physics~{\bf 919}, 166 (2016).

        \bibitem{On2024} M.~B.~On, F.~Ashtiani, D.~Sanchez-Jacome, D.~Perez-Lopez, S.~J.~B.~Yoo, and A.~Blanco-Redondo, Nat.~Commun.~{\bf 15}, 629 (2024).

        \bibitem{Liang2023} C.~Liang, Y.~Liu, F.~Li, S.~Leung, Y.~Poo, and J.~Jiang, Phys.~Rev.~Applied~{\bf 20}, 034028 (2023).

        \bibitem{Yu2022} Z.~Yu, H.~Lin, R.~Zhou, Z.~Li, Z.~Mao, K.~Peng, Y.~Liu, and X.~Shi, J.~Appl.~Phys.~{\bf 132}, 163101 (2022).

        \bibitem{Roberts2022} N.~Roberts, G.~Baardink, J.~Nunn, P.~J.~Mosley, and A.~Souslov, Sci.~Adv.~{\bf 8}, add3522 (2022).

        \bibitem{Henriques2022} J.~C.~G.~Henriques, T.~G.~Rappoport, Y.~V.~Bludov, M.~I.~Vasilevskiy, and N.~M.~R.~Peres, Phys.~Rev.~A~{\bf 101}, 043811 (2020).

        \bibitem{Xu2022} X.~Xu, Y.~Zhao, H.~Wang, A.~Chen, and Y.~Liu, Front.~Phys.~{\bf 9}, 813801 (2022).

        \bibitem{Aravena2022} G.~Cáceres-Aravena, B.~Real, D.~Guzmán-Silva, A.~Amo, L.~E.~F.~F.~Torres, and R.~A.~Vicencio, Phys.~Rev.~{\bf 4}, 013185 (2022).

        \bibitem{Upadhyay2024} V.~Upadhyay, M.~T.~Naseem, Ö.~E.~Müstecaplıoğlu, and R.~Marathe, New~J.~Phys.~{\bf 26}, 013014 (2024).

        \bibitem{Hao2022} H.~Hao, S.~Han, Y.~Yang, D.~Liu, H.~Xue, G.~Liu, Z.~Cheng, B.~Zhang, and Y.~Luo, Adv.~Mater.~{\bf 34}, 31 2202257 (2022).

        \bibitem{Xia2023} S.~Xia, D.~Zhang, X.~Zhai, L.~Wang, and S.~Wen, Appl.~Phys.~Lett.~{\bf 123}, 101102 (2023).

        \bibitem{Smith2021} T.~B.~Smith, C.~Kocabas, and A.~Principi, J.~Phys.:~Condens.~Matter~{\bf 33}, 265003 (2021).

        \bibitem{Guan2021} Y.~G., Z.~Jiang, and S.~Haas, Phys.~Rev.~B~{\bf 104}, 125425 (2021).

        \bibitem{Pocock2018} S.~R.~Pocock, X.~Xiao, P.~A.~Huidobro, and V.~Giannini, ACS~Photonics~{\bf 5}(6), 2271–2279 (2018).

        \bibitem{Downing2017} C.~A.~Downing and G.~Weick, Phys.~Rev.~B~{\bf 95}, 125426 (2017).

        \bibitem{Dong2024} Z.~Dong, H.~Li, T.~Wan, Q.~Liang, Z.~Yang, and B.~Yan, Nat.~Photonics~{\bf 18}, 68–73 (2024).

        \bibitem{LZhou2022} L.~Zhou, H.~Li, W.~Yi, and X.~Cui, Commun.~Phys.~{\bf 5}, 252 (2022).

        \bibitem{Jiang2022} J.~Jiang, J.~Zhang, F.~Mei, Z.~Ji, Y.~Hu, J.~Ma, L.~Xiao, and S.~Jia, Phys.~Rev.~A~{\bf 106}, 023318 (2022).

        \bibitem{Cooper2019} N.~R.~Cooper, J.~Dalibard, and I.~B.~Spielman, Rev.~Mod.~Phys.~{\bf 91}, 015005 (2019).

        \bibitem{He2018} Y.~He, K.~Wright, S.~Kouachi, and C.~Chien, Phys.~Rev.~A~{\bf 97}, 023618 (2018).
        
        \bibitem{P2022} P.~Wei, J.~Ni, X.~Zheng, D.~Liu, and L.~Zou, J.~Phys.~Condens.~Matter~{\bf 34}, 495801 (2022).

        \bibitem{YLi2021} Y.~Li and R.~Cheng, Phys.~Rev.~B~{\bf 103}, 014407 (2021).

        \bibitem{Guo2024} T.~Guo, B.~Assouar, B.~Vincen, A.~Merkel, J.~Appl.~Phys.~{\bf 135}, 043102 (2024).

        \bibitem{XYang2023} X.~Yang, H.~Jia, P.~Zhang, S.~Wang, Y.~Yang, Y.~Yang, and X.~Li, J.~Appl.~Phys.~{\bf 133}, 195104 (2023).

        \bibitem{QLi2023} Q.~Li, X.~Xiang, L.~Wang, Y.~Huang, and X.~Wu, Appl.~Phys.~Lett.~{\bf 122}, 191704 (2023).

        \bibitem{Coutant2022} A.~Coutant, V.~Achilleos, O.~Richoux, G.~Theocharis, and V.~Pagneux, J.~Acoust.~Soc.~Am.~{\bf 151}, 3626–3632 (2022).

        \bibitem{Peng2018} Y.~Peng, Z.~Geng, and X.~Zhu, J.~Appl.~Phys.~{\bf 123}, 091716 (2018).

        \bibitem{Banerjee2023} D.~Banerjee, J.~Thomas, A.~Nocera, and S.~Johnston, Phys.~Rev.~B {\bf 107}, 235113 (2023).

        \bibitem{Rosenberg2022} P.~Rosenberg and E.~Manousakis, Phys.~Rev.~B {\bf 106}, 054511 (2022).

        \bibitem{Wang2021} Z.~Wang, F.~Xu, L.~Li, D.~Xu, W.~Chen, and B.~Wang, Phys.~Rev.~B {\bf 103}, 134507 (2021).


        \bibitem{Rosenberg2021} P.~Rosenberg and E.~Manousakis, Phys.~Rev.~B {\bf 104}, 134511 (2021).
        
        \bibitem{Zhao2023} X.~Zhao, Y.~Xing, J.~Cao, S.~Liu, W.~Cui, and H.~Wang, npj Quantum Information {\bf 9}, 59 (2023).

        \bibitem{Zheng2022} L.~Zheng, X.~Yi, and H.~Wang, Phys.~Rev.~Applied {\bf 18}, 054037 (2022).

        \bibitem{Chen2022} J.~Chen, C.~Wu, J.~Fan, and G.~Chen, Chinese Phys.~B {\bf 31}, 088501 (2022).

        \bibitem{Navarro-Labastida2022} L.~A.~Navarro-Labastida, F.~A.~Domınguez-Serna, and F.~Rojas, Revista Mexicana de Fısica {\bf 68}, 031404 (2022).

        \bibitem{Fromholz2020} P.~Fromholz, G.~Magnifico, V.~Vitale, T.~Mendes-Santos, and M.~Dalmonte, Phys.~Rev.~B {\bf 101}, 085136 (2020).

        \bibitem{Micallo2020} T.~Micallo, V.~Vitale, M.~Dalmonte, and P.~Fromholz, SciPost Phys.~Core {\bf 3}, 012 (2020).

        \bibitem{Zurita2023} J.~Zurita, C.~E.~Creffield, and G.~Platero, Quantum {\bf 7}, 1043 (2023).

        \bibitem{Chang2023} Y.~Chang, J.~Xue, Y.~Han, X.~Wang, and H.~Li, Phys.~Rev.~A {\bf 108}, 062409 (2023).

        \bibitem{Zheng2023} L.~Zheng, H.~Wang, and X.~Yi, New J.~Phys.~{\bf 25}, 113003 (2023).

        \bibitem{CWang2022} C.~Wang, L.~Li, J.~Gong, and Y.~Liu, Phys.~Rev.~A {\bf 106}, 052411 (2022).

        \bibitem{Palaiodimopoulos2021} N.~E.~Palaiodimopoulos, I.~Brouzos, F.~K.~Diakonos, and G.~Theocharis, Phys.~Rev.~A {\bf 103}, 052409 (2021).

        \bibitem{D'Angelis2020} F.~M.~D'Angelis, F.~A.~Pinheiro, D.~Guéry-Odelin, S.~Longhi, and François~Impens, Phys.~Rev.~Research {\bf 2}, 033475 (2020).

        \bibitem{Tan2020} S.~Tan, R.~W.~Bomantara, and J.~Gong, Phys.~Rev.~A {\bf 102}, 022608 (2020).

        \bibitem{Mei2018} F.~Mei, G.~Chen, L.~Tian, S.~Zhu, and S.~Jia, Phys.~Rev.~A {\bf 98}, 012331 (2018).

        \bibitem{Cinnirella2024} E.~G.~Cinnirella, A.~Nava, G.~Campagnano, and D.~Giuliano, Phys.~Rev.~B {\bf 109}, 035114 (2024).

        \bibitem{Bera2023} M.~L.~Bera, J.~O.~de Almeida, M.~Dziurawiec, M.~Płodzień, M.~M.~Maśka, M.~Lewenstein, T.~Grass, and U.~Bhattacharya, Phys.~Rev.~B {\bf 108}, 214104 (2023).

        \bibitem{Dias2022} R.~G.~Dias and A.~M.~Marques, Phys.~Rev.~B {\bf 105}, 035102 (2022).

        \bibitem{Qi2021} L.~Qi, Y.~Yan, Y.~Xing, X.~Zhao, S.~Liu, W.~Cui, X.~Han, S.~Zhang, and Hong-Fu~Wang, Phys.~Rev.~Research {\bf 3}, 023037 (2021).

        \bibitem{Pérez-González2019} B.~Pérez-González, M.~Bello, Á.~Gómez-León, and G.~Platero, Phys.~Rev.~B {\bf 99}, 035146 (2019).

        \bibitem{Malakar2023} R.~K.~Malakar and A.~K.~Ghosh, J.~Phys.: Condens.~Matter {\bf 35}, 335401 (2023).
        
        \bibitem{Zhang2017} S.~Zhang and Q.~Zhou, Phys.~Rev. A {\bf 95}, 061601(R) (2017). 
        
        \bibitem{Koor2022} K.~Koor, R.~W.~Bomantara, and L.~C.~Kwek, Phys.~Rev.~B {\bf 106}, 195122 (2022).

        \bibitem{Feng2022} C.~Feng, B.~Xing, D.~Poletti, R.~Scalettar, and G.~Batrouni, Phys.~Rev.~B {\bf 106}, L081114 (2022).

        \bibitem{Yu2020} X.~Yu, L.~Jiang, Y.~Quan, T.~Wu, Y.~Chen, L.~Zou, and J.~Wu, Phys.~Rev.~B {\bf 101}, 045422 (2020).

        \bibitem{Melo2023} P.~B.~Melo, S.~A.~S.~Júnior, W.~Chen, R.~Mondaini, and T.~Paiva, Phys.~Rev.~B {\bf 108}, 195151 (2023).

        \bibitem{Jezequel2022} L.~Jezequel and P.~Delplace, Phys.~Rev.~B {\bf 105}, 035410 (2022).

        \bibitem{Y.Ma2021} Y.-P.~Ma and H.~Susanto, Phys.~Rev.~E {\bf 104}, 054206 (2021).

        \bibitem{Tuloup2020} T.~Tuloup, R.~W.~Bomantara, C.~H.~Lee, J.~Gong, Phys. Rev.~B~{\bf 102}, 115411 (2020).

        \bibitem{Jangjan2024} M.~Jangjan, L.~Li, L.~E.~F.~F.~Torres, and M.~V.~Hosseini, Phys.~Rev.~B {\bf 109}, 205142 (2024).

        \bibitem{Yao2018} S.~Yao and Z.~Wang, Phys.~Rev.~Lett.~{\bf 121}, 086803 (2018). 
       
        \bibitem{Lee2016} T.~E.~Lee, Phys.~Rev.~Lett.~{\bf 116}, 133903 (2016). 

        \bibitem{Lieu2018} S.~Lieu, Phys.~Rev.~B~{\bf~97}, 045106 (2018).

        \bibitem{Wu2021} H.~C.~Wu, L.~Jin, and Z.~Song, Phys.~Rev.~B~{\bf~103}, 235110 (2021).

        \bibitem{Halder2023} D.~Halder, S.~Ganguly, and S.~Basu, J.~Phys.: Condens.~Matter~{\bf~35}, 105901 (2023).

        \bibitem{Okuma2023} N.~Okuma and M.~Sato, Annu.~Rev.~Condens.~Matter~Phys.~{\bf 14}, 83-107 (2023). 

        \bibitem{Asboth2014} J.~K.~Asboth, B.~Tarasinski, P.~Delplace, Phys.~Rev.~B~{\bf 90}, 125143 (2014).

        \bibitem{Bomantara2019} R.~W.~Bomantara, L.~Zhou, J.~Pan, J.~Gong, Phys.~Rev.~B~{\bf 99}, 045441 (2019).

        \bibitem{Qiao2023} Q.~Qiao, L.~Wang, G.~Li, X.~Chen, and L.~Yuan, Nanophotonics vol. {\bf 12},  pp. 3807-3815 (2023).

        \bibitem{Wu2020} H.~Wu and J.~An, Phys.~Rev.~B~{\bf 102}, 041119(R) (2020).

        \bibitem{Pan2020} Y.~Pan and B.~Wang, Phys.~Rev.~Research~{\bf 2}, 043239 (2020).
        
        \bibitem{Jangjan2022} M.~Jangjan, L.~E.~F.~F.~Torres, and M.~V.~Hosseini, Phys.~Rev.~B~{\bf 106}, (224306) 2022.

        \bibitem{Verma2024} S.~Verma and T.~K.~Ghosh. “Emergent SU(3) topological system in a trimer SSH model.” (2024).

        \bibitem{Anastasiadis2022} L.~A.~Anastasiadis, G.~Styliaris, R.~Chaunsali, G.~ Theocharis, and F.~K.~Diakonos, Phys.~Rev.~B~{\bf 106}, 085109 (2022).
        
        \bibitem{Alvarez2019} V.~M.~M.~Alvarez and M.~D.~Coutinho-Filho, Phys.~Rev.~A~{\bf 99}, 013833 (2019).

        \bibitem{Du2024} T.~Du, Y.~Li, H.~Lu, and H.~Zhang,  New J. Phys.~{\bf 26}, 023044 (2024).

        \bibitem{Zhou2023} X.~Zhou, J.~Pan, and S.~Jia, Phys.~Rev.~B~{\bf 107}, 054105 (2023).

        \bibitem{Marques2020} A.~M.~Marques and R.~G.~Dias, J. Phys. A: Math. Theor. {\bf 53}, 075303  (2020).
        
        \bibitem{Xie2019} D.~Xie, W.~Gou, T.~Xiao, B. Gadway, and B.~Yan, npj Quantum Information {\bf 5}, 55 (2019).

        \bibitem{Deng2022} J.~Deng, H.~Dong, C.~Zhang, Y.~Wu, J.~Yuan, X.~Zhu, F.~Jin, H.~Li, Z.~Wang, H.~Cai, C.~Song, H.~Wang, J.~Q.~You, and D.~Wang, Science~{\bf 378}, 966-971 (2022).

        \bibitem{Youssefi2022} A.~Youssefi, S.~Kono, A.~Bancora, M.~Chegnizadeh, J.~Pan, T.~Vovk, and T.~J.~Kippenberg, Nature~{\bf 612}, 666–672 (2022).

        \bibitem{Cai2019} W.~Cai, J.~Han, F.~Mei, Y.~Xu, Y.~Ma, X.~Li, H.~Wang, Y.~P.~Song, Z.~Xue, Z.~Yin, S.~Jia, and L.~Sun, Phys.~Rev.`Lett.~{\bf 123}, 080501 (2019).
        
        \bibitem{Cheng2019} Q.~Cheng, Y.~Pan, H.~Wang, C.~Zhang, D.~Yu, A.~Gover, H.~Zhang, T.~Li, L.~Zhou, and S.~Zhu, Phys.~Rev.~Lett.~{\bf 122}, 173901 (2019).

        \bibitem{Coutant2021} A.~Coutant, A.~Sivadon, L.~Zheng, V.~Achilleos, O.~Richoux, G.~Theocharis, V.~Pagneux, Phys.~Rev.~B~{\bf 103}, 224309 (2021).

        \bibitem{Yang2024} Y.~Yang, R.~J.~Chapman, B.~Haylock, F.~Lenzini, Y.~N.~Joglekar, M.~Lobino, and A.~Peruzzo, Nat. Commun.~{\bf 15}, 50 (2024). 


        \bibitem{Sougleridis2024} I.~I.~Sougleridis, A.~Anastasiadis, O.~Richoux, V.~Achilleos, G.~Theocharis, V.~Pagneux, F.~K.~Diakonos, ``Existence and characterization of edge states in an acoustic trimer Su-Schrieffer-Heeger model'' (2024).

        \bibitem{Savelev2020} R.~S.~Savelev and M.~A.~Gorlach, Phys.~Rev.~B~{\bf 102}, 161112(R) (2020).

        \bibitem{XinLi2018} X.~Li, Y.~Meng,  X.~Wu,  S.~Yan,  Y.~Huang, S.~Wang, and W.~Wen, Appl. Phys.~Lett.~{\bf 113}, 203501 (2018).

        \bibitem{AZ97} A.~Altland and M.~R.~Zirnbauer, Phys.~Rev.~B~{\bf 55}, 1142 (1997).

        \bibitem{O2017} O.~V.~Ogloblya, H.~M.~Kuznietsova, and Y.~M.~Strzhemechny, Physica~B~Condens.~Matter~{\bf 504}, 96 (2017).  

        \bibitem{Huang11} L.-T.~Huang and D.-H.~Lee, Phys.~Rev.~B~{\bf 84}, 193106 (2011)
        \bibitem{Z.Cheng2023} Z.~Cheng, R.~W.~Bomantara, H.~Xue, W.~Zhu, J.~Gong, and B.~Zhang, Phys.~Rev.~Lett.~{\bf 129}, 254301 (2023).


   \end{thebibliography}
\end{document}